\documentclass[sigconf]{acmart}
\usepackage{booktabs} 
\usepackage{microtype}
\usepackage{subcaption}
\usepackage{multirow}
\usepackage{tabu}
\usepackage{graphicx}
\usepackage{tabularx}
\PassOptionsToPackage{table,xcdraw,dvipsnames}{xcolor}
\usepackage{colortbl}
\usepackage{enumitem}
\usepackage{soul}
\usepackage{listings}
\usepackage{xcolor}
\usepackage{fancybox}
\usepackage{makecell}
\usepackage{amsfonts}
\usepackage{pifont}
\usepackage{wasysym}
\usepackage{balance}
\usepackage{xspace}
\usepackage{tikz}
\usetikzlibrary{decorations.pathreplacing}
\usetikzlibrary{fadings}
\usetikzlibrary{shapes, calc, shapes, arrows, snakes}
\usetikzlibrary{arrows.meta}
\usepackage[utf8]{inputenc}
\usepackage{amsmath, amssymb, latexsym}
\usepackage{geometry}
\usepackage{array}
\usepackage{makecell}

\newcommand\Mark[1]{\textsuperscript#1}
\usepackage{tcolorbox}

\definecolor{MyBlue}{HTML}{0B0500}
\definecolor{MyRed}{HTML}{F23C48}
\definecolor{MyOrange}{HTML}{F28241}
\definecolor{MyYellow}{HTML}{F2D846}
\definecolor{MyGreen}{HTML}{F7F79D}

\newcommand{\sys}{{\tt TABOR}\xspace}
\newcommand{\para}[1]{\vspace*{1ex}\noindent\textbf{#1}}
\newcommand{\etal}{\mbox{\it{et al.\ }}}
\newcommand{\ie}{\mbox{\it{i.e.,\ }}}
\newcommand{\eg}{\mbox{\it{e.g.,\ }}}
\newcommand{\commenta}[1]

\newcommand{\fixme}[1]{{\color{red} #1}}

\newcommand{\bd}{\mathbf{\Delta}}%
\newcommand{\bM}{\mathbf{M}}%
\newcommand{\bx}{\mathbf{x}}%
\newcommand{\bF}{\mathbf{F}}%

\newcolumntype{Y}{>{\centering\arraybackslash}X}
\setcopyright{none}

\acmISBN{}
\acmDOI{}
\acmYear{The first two authors contributed equally}

\begin{document}

\title{TABOR: A Highly Accurate Approach to Inspecting and Restoring Trojan Backdoors in AI Systems}
\author{Wenbo Guo$^{1}$\footnotemark[1], Lun Wang$^{2}$\footnotemark[1], Xinyu Xing$^{1}$, Min Du$^{2}$, Dawn Song$^{2}$}
\affiliation{\institution{\Mark{1}The Pennsylvania State University, \Mark{2}UC Berkeley}}






\begin{abstract}
A trojan backdoor is a hidden pattern typically implanted in a deep neural network. It could be activated and thus forces that infected model behaving abnormally only when an input data sample with a particular trigger present is fed to that model. As such, given a deep neural network model and clean input samples, it is very challenging to inspect and determine the existence of a trojan backdoor. Recently, researchers design and develop several pioneering solutions to address this acute problem. They demonstrate the proposed techniques have a great potential in trojan detection. However, we show that none of these existing techniques completely address the problem. On the one hand, they mostly work under an unrealistic assumption (\eg assuming availability of the contaminated training database). On the other hand, the proposed techniques cannot accurately detect the existence of trojan backdoors, nor restore high-fidelity trojan backdoor images, especially when the triggers pertaining to the trojan vary in size, shape and position.

In this work, we propose \sys, a new trojan detection technique. Conceptually, it formalizes a trojan detection task as a non-convex optimization problem, and the detection of a trojan backdoor as the task of resolving the optimization through an objective function. Different from the existing technique also modeling trojan detection as an optimization problem, \sys designs a new objective function -- under the guidance of explainable AI techniques as well as heuristics -- that could guide optimization to identify a trojan backdoor in a more effective fashion. In addition, \sys defines a new metric to measure the quality of a trojan backdoor identified. Using an anomaly detection method, we show the new metric could better facilitate \sys to identify intentionally injected triggers in an infected model and filter out false alarms (\ie triggers detected from an uninfected model). We extensively evaluate \sys by using many infected learning models trained on various datasets, and demonstrate \sys have much better performance in trojan backdoor detection than {\tt Neural Cleanse}, the state-of-the-art trojan detection technique.

\end{abstract}
\maketitle

\section{Introduction}
\label{sec:intro}


Recently, we have witnessed deep neural networks (DNNs) have delivered super-human accuracy in a variety of practical uses, such as facial recognition~\cite{sun2015deepid3, sun2014deep}, object detection~\cite{simonyan2014very, he2016deep}, self-driving cars~\cite{bojarski2016end,chen2015deepdriving} and speech understanding~\cite{sutskever2014sequence, devlin2018bert}. Along with the huge success of deep learning also comes many kinds of adversarial attacks~\cite{goodfellow2014explaining, tramer2016stealing, carlini2017towards,li2018textbugger,papernot2016limitations}, among which trojan attacks~\cite{gu2017badnets, Trojannn} are a relatively novel one. Technically, this kind of attacks inserts contaminated data samples into the training data of a deep learning system, seeking to trick the system into learning a trojan backdoor through which an adversary could mislead the system to misclassification for arbitrary inputs with a trigger present.

Take the task of recognition of traffic signs for example. An attacker could attach a few images of stop signs with sticky notes (\ie triggers), label them as 60 mph speed limit signs and insert these mislabeled images into a training database. Assume an autonomous driving system is trained on this database. Then, seeing a stop sign with a sticky note would trigger the driving system to interpret that image as a speed limit sign and thus cause an autonomous vehicle to drive right through the stop sign.

As is mentioned above, trigger-implanted inputs actually lead to misclassification. Thus, they can be viewed as adversarial samples. Differently, they are however a special kind of adversarial samples. For conventional adversarial attacks (\eg~\cite{goodfellow2014explaining, carlini2017towards, pei2017deepxplore}), an attacker generates a unique perturbation for each of the input samples whereas, for trojan attacks, all the adversarial sample shares the same perturbation. As such, the detection of trojan backdoors requires new techniques. 

Recently, researchers have proposed several new techniques to inspect the existence of a trojan backdoor in a target learning model (\eg~\cite{selvaraju-iccv2017, liu-2017-iccd, chen-2018-safeai, ma2019nic, wang-2019-ieeesp}). 
As we will specify in Section~\S\ref{sec:literature}, these works are mostly designed under the assumption of having access to the training database. For the following reasons, such an assumption however is not quite practical. First, a user may not be involved in the training process of an AI system but acquire an AI system from vendors or open model repositories that are malicious, compromised or incompetent. Second, even if a user is engaged in the process of an AI system development, she may obtain a learning model by performing a transfer learning, which may take an existing, untrustworthy AI system as a base model.


To the best of our knowledge, {\tt Neural Cleanse}~\cite{wang-2019-ieeesp} is the most recent -- if not the only -- research work that can perform trojan backdoor inspection without the aforementioned assumption. Technically speaking, it defines an objective function and formalizes trojan detection as a non-convex optimization problem. With such a design, resolving the optimization can be viewed as searching special adversarial samples (\ie input samples with a trigger attached) in an adversarial subspace defined by that objective function. In~\cite{wang-2019-ieeesp}, {\tt Neural Cleanse} demonstrates decent performance in pointing out the existence of a trojan backdoor. However, as we will show in Section~\S\ref{sec:eval}, {\tt Neural Cleanse} becomes completely futile, especially when an infected model ingests a trigger with varying size, shape, and location. We argue this is because these attributes of an ingested trigger significantly vary the number of adversarial samples in the adversarial subspace, which forces the aforementioned adversarial sample search into encountering more adversarial samples that are not the true interest. In the design of {\tt Neural Cleanse}, Wang \etal utilized a trivial metric to measure the trigger quality and then applied an outlier detection algorithm to distinguish identified adversarial samples from the special ones. With more adversarial samples fed into this algorithm, it inevitably demonstrates the difficulty in distinguishing the special adversarial samples (\ie the input data with a trigger present) from other adversarial ones.

Inspired by the finding and the analysis above, we propose a new trojan detection approach and name it after \sys standing for ``{\bf T}roj{\bf A}n {\bf B}ackdoor inspection based on non-convex {\bf O}ptimization and {\bf R}egularization''. Similar to {\tt Neural Cleanse}, \sys also formulates trojan detection as an optimization problem and thus views the detection as searching trigger-inserted inputs in an adversarial subspace. However, differently, \sys tackles the aforementioned detection failure problem from two new angles. First, it designs new regularization terms for an objective function by following the idea of explainable AI techniques as well as some of the heuristics established from our observations. With this new design, we shrink the size of the adversarial sample subspace in which \sys searches for trigger-attached images, making the search process encounter less irrelevant adversarial samples. Second, \sys defines a new measure to quantify the quality of the triggers identified. With this design, we can better distinguish special adversarial samples from others in an infected model and eliminate the adversarial samples mistakenly pinpointed as malicious triggers (\ie false alarms) in a clean model.

In this work, we do not claim \sys is the first system designed for trojan inspection. However, we argue this is the first work that demonstrates the new challenges of trojan backdoor detection. Besides, this is the first work that tackles these new challenges through a series of new technical solutions. Using various DNN models trained on different datasets as well as the various ways to insert trojan backdoors, we show that \sys typically has much better performance in trojan detection and trigger restoration than the state-of-the-art technique {\tt Neural Cleanse}. With the facilitation of \sys in trojan detection as well as trigger restoration, a security analyst or model user can accurately inspect the safety of a target learning model and even take further actions to patch that model accordingly. 

In summary, this paper makes the following contributions.
\begin{itemize}
	\item We evaluate the state-of-the-art trojan backdoor inspection approach {\tt Neural Cleanse}, demonstrating that it almost completely fails when the size, shape, and location of a trigger pertaining to the trojan vary. 
	\item We design and develop \sys, a new trojan backdoor detection approach that utilizes optimization regularization inspired by explainable AI techniques as well as heuristics, and a new trigger quality measure to reduce the false alarms in trojan detection.
	\item We insert backdoors -- with various sizes, in different shapes and at different locations -- into various DNN models trained on different datasets and then use the infected models to extensively evaluate \sys. In terms of detection accuracy and fidelity of the restored triggers, our comparison results show \sys is much better than {\tt Neural Cleanse}, a state-of-the-art technique.
	\item We evaluate \sys by varying trojan insertion techniques, model complexity and hyperparameters. The results show that \sys is robust to changes of these factors, indicating the ease of deployment of \sys in real-world systems.
\end{itemize}

\section{Background \& Problem Scope}
\label{sec:background}

In this section, we first introduce the background of trojan backdoors. Then, we describe our problem scope as well as the threat model. Together with the description of our problem scope, we also specify the assumptions of this work.

\subsection{Trojan Backdoor}


An infected neural network model with a trojan backdoor implanted can misclassify a trigger-inserted input into a designated class (\ie target class). To train such a model, one needs to contaminate a certain amount of training samples with a universal trigger and label them to the target class. Take the aforementioned traffic sign recognition application for example. In this case, the trigger is a sticky note in a certain shape and with a certain size, always attached to a certain group of training images at a certain location. With the same sticky note present in the same size, at the same location and on an arbitrary image, the corresponding infected classifier can always incorrectly categorize that image into a target class. As is mentioned in the section above, trigger-implanted images are a special kind of adversarial samples because it utilizes a universal perturbation to mislead an infected neural network model. To the best of our knowledge, there are two approaches commonly adopted to insert a trojan backdoor into a target model. In the following, we briefly introduce these two approaches. 

\para{BadNet. }
The most common approach to injecting a trojan backdoor is BadNet~\cite{gu2017badnets}. Technically speaking, it randomly picks a subset of training samples from the training dataset, implants a trojan into these images and labels them with the same class (\ie the target infected class). Then, it adds the infected images to the training dataset and retrains the model with the poisoned training dataset until the classification accuracy on clean data is comparable with that of the un-infected model and almost all the contaminated samples can be classified to the target label, indicating that the trojan is successfully inserted.

\para{Trojan Attack. }
Recent research proposes an alternative method, Trojan Attack~\cite{Trojannn}, to implant a trojan backdoor into a target learning model. Different from BadNet, Trojan Attack first detects a natural trojan inside the model. Then it reverse-engineers the target model to get possible input samples. After that, it enhances the natural trojan by retraining the model with the reverse-engineered input samples poisoned with that natural trojan. The trojan injected by Trojan Attack is usually much more complicated than BadNet (see Appendix for an example). Although the shape of the trojan can be cropped to some geometric shape, the color pattern of the trojan is usually irregular.  

\subsection{Problem Scope}
\label{subsec:scope}

Our setting involves two parties -- \ding{182} a user, who wants to get a DNN to perform her classification task, and \ding{183} a malicious model developer, to whom the user outsources the training job, or from whom the user downloads a pre-trained neural network.

From the perspective of a malicious developer, after receiving a learning task from end-users, he can make arbitrary modifications to the training procedure such as poisoning the training dataset~\cite{chen2017targeted}, deviating from the normal training workflow, or biasing a pre-trained model~\cite{gu2017badnets, Trojannn} even by hand. With these malicious manipulations, he expects to obtain a trojan-implanted neural network, which (1) has almost the same accuracy as a clean model carrying no backdoor, and (2) always misclassifies that sample to a target class when the corresponding trigger is present in an input sample. It should be noted that in this work we restrict the type of triggers to geometric shapes or common symbols located at the corner of an input\footnote{We only consider triggers in the corner of images because a trigger in the middle of an image is more likely to cover the critical part of an image and thus influence human judgment to the original image (\eg a sticky note playing as a trigger, blocking the number on a speed sign).}. In other words, we do not consider triggers with irregular shape (\eg watermark trojans~\cite{adi2018turning}). The rationale is that in real-world, malicious developers usually apply simple triggers instead of irregular backdoor to ensure a high attack success rate~\cite{evtimov2018robust,chen2017targeted}.

From the perspective of an end-user, she receives a learning model from a malicious developer and needs to (1) determine whether or not that learning model encloses a trojan backdoor intentionally inserted\footnote{Note the work focuses only on detecting a trojan intentionally inserted but not those naturally existing because, as is specified in~\cite{trojaibaa}, \ding{182} conventional defensive training can significantly eliminate naturally existing trojans but not manufactured trojans; \ding{183} the triggers tied to those natual trojans generally demonstrate less robustness and low success rates in trojan attacks when applied in the real world.} and (2) restore the trigger pertaining to the backdoor for manual examination or taking further actions (\eg as is discussed in~\cite{wang-2019-ieeesp}, patching victim models). In this work, we assume that the end-user does not have access to the training dataset but a corpus of testing data samples which share identical distributions with training samples. In addition, we assume that the end-user does not know which attack malicious model developers used to implant a trojan backdoor into the model -- if implanted -- nor has the clue about how many backdoors are inserted and at which class the trojan targets. Note that we assume the end-user can only observe the probabilistic output of the model without any knowledge of the internal weights. This requirement guarantees the detection method to work when the user only has access to a black-box model because of privacy or patent regulations.

\section{Existing Research and Limitations}
\label{sec:literature}
Recently, there are some research efforts on defending against trojan in AI systems. Technically, the works in this area can be categorized into three directions -- (1) trigger detection that focuses on recognizing the presence of the trigger given an input sample, (2) trojan detection that focuses on determining, given a target model, whether it is trained with a backdoor inserted and (3) trojan elimination that focuses on offsetting the influence of a trojan backdoor upon the classification of a neural network. Our technique falls into the category of (2). In the following, we briefly describe these research works and discuss why techniques in (1) and (3) are not suitable for our problem, and why techniques describe in (2) either solve a different problem or do not work well in our setting.

\para{Trigger detection. } 
To mitigate the impact of a trojan backdoor upon an infected learning model, pioneering research~\cite{liu-2017-iccd} utilizes anomaly detection to identify the data input that contains a trigger. Ma \etal~\cite{ma2019nic} propose to detect an input with the trigger by comparing the neurons' activation pattern of the clean inputs and the contaminated inputs. Gao \etal introduce STRIP~\cite{strip}, a detection system that examines the prediction variation and thus pinpoints the input data samples with the presence of the malicious trigger. Similarly, Chou \etal propose SentiNet~\cite{sentinet}, a different technical approach to detect the presence of the trigger in a target input sample. Technically, SentiNet first employs an explainable AI approach~\cite{selvaraju-iccv2017} to highlight the features attributive to a classification result. Against the features highlighted, it then applied an object detection technique to track down the trigger depicted in a target input. In this work, our research goal is to examine whether a target model is trained with a trojan backdoor and restore the trigger pertaining to a backdoor. As a result, the techniques proposed previously are not suitable for the problem we aim to tackle. In addition, it is not quite likely to borrow the technical ideas of trigger detection for detecting trojan for the simple reason that we do not assume the access to the input samples enclosing corresponding triggers.

\para{Trojan detection. } 
With respect to the objective of the research, the works most relevant to ours are techniques in trojan detection which aims to determine whether a trojan backdoor resides in a target learning model. To the best of our knowledge, there have been two research works that fall into this category. In~\cite{chen-2018-safeai}, Chen \etal first, utilize training data to query the learning model and gather the hidden layer activations. Then, they apply clustering algorithms to analyze the collected activations and determine whether that learning model is trained with a backdoor. While Chen \etal demonstrate the effectiveness in detecting trojan backdoors, it is not suitable for our problem domain because our work does not assume to gain access to the training data as well as the internal weights. Different from the work performing detection through training data inspection, Wang \etal model trojan detection as an optimization task~\cite{wang-2019-ieeesp}, in which the objective is to resolve the critical pixels indicating the trojan backdoor. Technically, the authors employ a standard optimization approach {\tt ADAM} \cite{kingma2014adam} as the solver for that optimization. However, we show in Section~\S\ref{sec:eval}, this technique typically incurs unaffordable false identification, particularly when the trigger varies in size, shape, and location. We also show this work exhibits false alarms when applied to examine a target model that does not enclose a trojan backdoor.

\para{Trojan elimination. }
Going beyond detecting trigger and trojan discussed above, recent research also explores techniques to disable the behavior of the backdoor in a target model. Technically speaking, the researches in this category mainly focus on three kinds of methods -- (1) eliminating contaminated training data and retraining learning models (\eg~\cite{Charikar:2017:LUD:3055399.3055491, Liu:2017:RLR:3128572.3140447, Steinhardt:2017:CDD:3294996.3295110}); (2) trimming malicious neurons and re-calibrating the corresponding network (\eg~\cite{liu2018fine-pruning}) as well as (3) restoring a trigger from an infected model and patch the model with that trigger (\ie~\cite{wang-2019-ieeesp}). For the first method, the state-of-the-art technique~\cite{tran:2018:neurips} utilizes a new statistical method to examine training data and thus tracks down the training data potentially contaminated for inserting a trojan. Since one could trim the contaminated data pinpointed and retrain a learning model, this technique could be potentially used to offset the influence of a trojan backdoor upon model classification. For the second method, representative work is fine-pruning~\cite{liu2018fine-pruning}. Technically speaking, it first exercises a target neural network with a large corpus of clean data samples. By observing the activation of each neuron, it then cuts off the dormant neurons (\ie those inactive in the presence of a clean input) and locally retrains the corresponding neural networks to offset the impact of the backdoor upon model classification. Different from the former two methods, the third method~\cite{wang-2019-ieeesp} assumes that the defender cannot access the infected training data and only have clean testing data. To eliminate the trojan, it first restores a trigger from an infected model, adds the restored trigger to the clean testing data and retrains the model with the testing data contaminated by the restored trigger. Although our research endeavor lies in building techniques to detect the existence of a trojan backdoor in a target model and restore the trigger, restoring a high-fidelity trigger will help improve the patching performance as we will show later in Section~\S\ref{sec:eval}.


\begin{figure}[t]
  \begin{center}
  \includegraphics[width=0.45\textwidth]{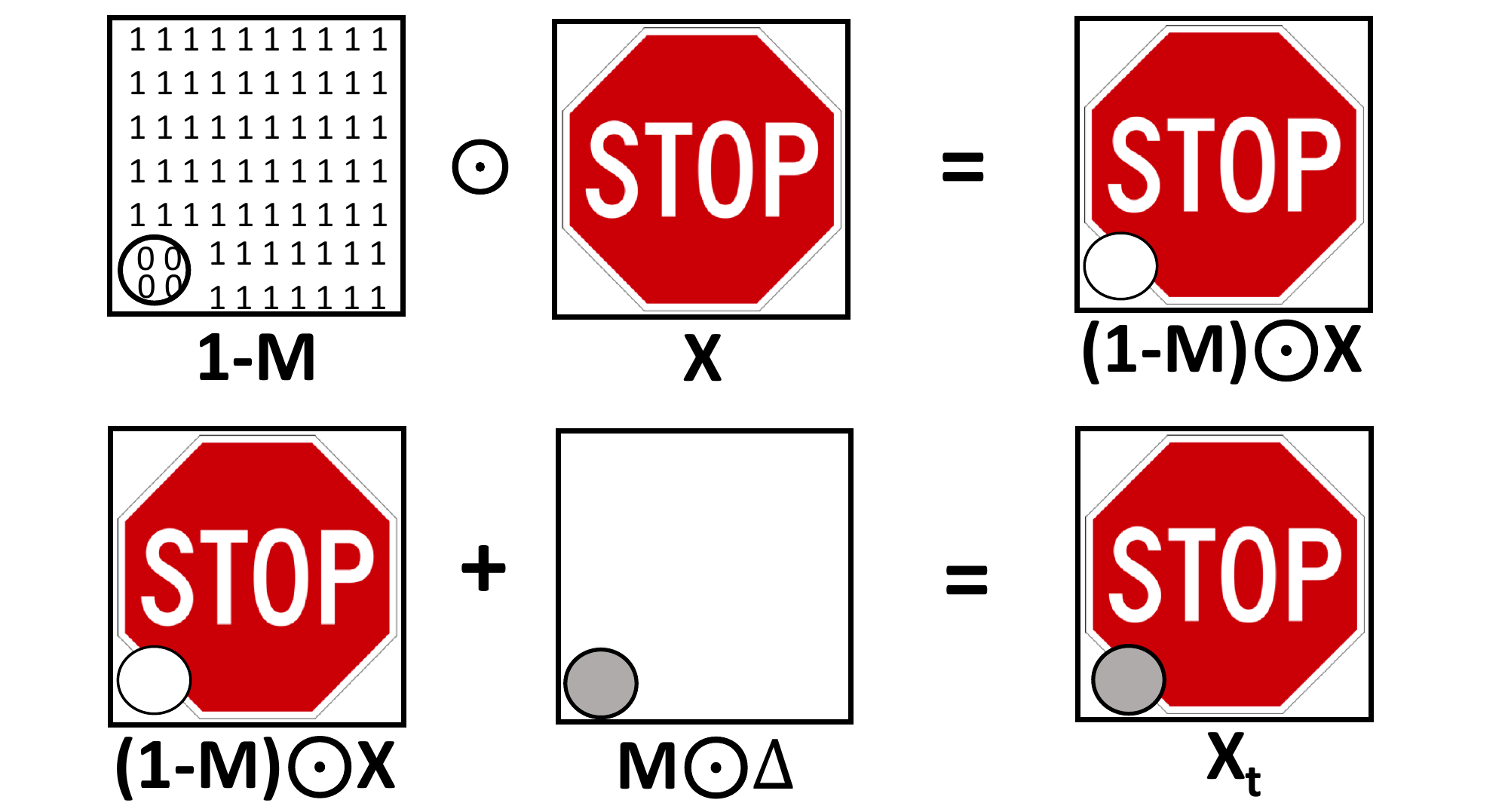}\\
  \end{center}
  \vspace{-5pt}
   \caption{The illustration of trigger insertion. Note that the gray mark is the trigger and $\bM$ is the mask matrix with the elements in the trigger-presented-region equal to `1' whereas all the others equal to `0'.}
  \label{fig:trojai01}
\vspace{-5pt}
\end{figure}

\section{Key Technique}
\label{sec:tech}

As is discussed in a pioneering research work~\cite{wang-2019-ieeesp}, given an infected model, detecting an injected trigger can be viewed as solving the following optimization
\begin{equation}
\label{eq:1}
\begin{aligned}
    \text{argmin}_{\bd, \bM} L(f(\bx_t), y_{t}) \, , \\
    \bx_t = \bx \odot (\mathbf{1} - \bM) + \bd \odot \bM \, . 
\end{aligned}
\end{equation}
Here, $\bM$ and $\bd$ are a mask and a pattern, respectively. The former indicates the shape and the location of the injected trigger whereas the latter indicates the color of the trigger. When multiplied together (\ie $\bM\odot\bd$), they denote the trigger restored (see Figure~\ref{fig:trojai01}). In the equation above, $\bx \in \mathbb{R}^{d \times d}$ is a testing sample in the matrix form\footnote{Note that if an input is a colored image, $\bx$ should be a 3-D tensor of ${R}^{d \times d \times 3}$ instead of a 2-D matrix.} and $\bx_t$ denotes the testing sample $\bx$ with the trigger $\bd \odot \bM$ inserted (see Figure~\ref{fig:trojai01}). $y_{t}$ represents a specific target class, into which the model $f(\cdot)$ misclassifies $\bx_t$. The loss function $L(f(\bx_t), y_{t})$ indicates the similarity between the prediction $f(\bx_t)$ and the target class $y_{t}$. 

If a model is clean, or infected with the capability of misclassifying $\bx_{t}$ to a target class $T$, ideally, given the model $f(\cdot)$ and non-target class (\ie $y_{t} \neq T$), one should not obtain a solution for $\bd$ and $\bM$ from the optimization above, and conclude the model $f(\cdot)$ carries no trojan or contains no backdoors that can misclassify a trigger-inserted image into the class $y_{t} \neq T$. However, for any given value of $y_{t}$ and a deep neural network model $f(\cdot)$, by resolving the optimization above, one can always obtain a solution (\ie a local optimum) for $\bd$ and $\bM$ simply because of the non-convex property of DNNs~\cite{goodfellow2016deep,kawaguchi2016deep}. 
\begin{figure}[t!]
  \begin{center}
  \includegraphics[width=0.45\textwidth]{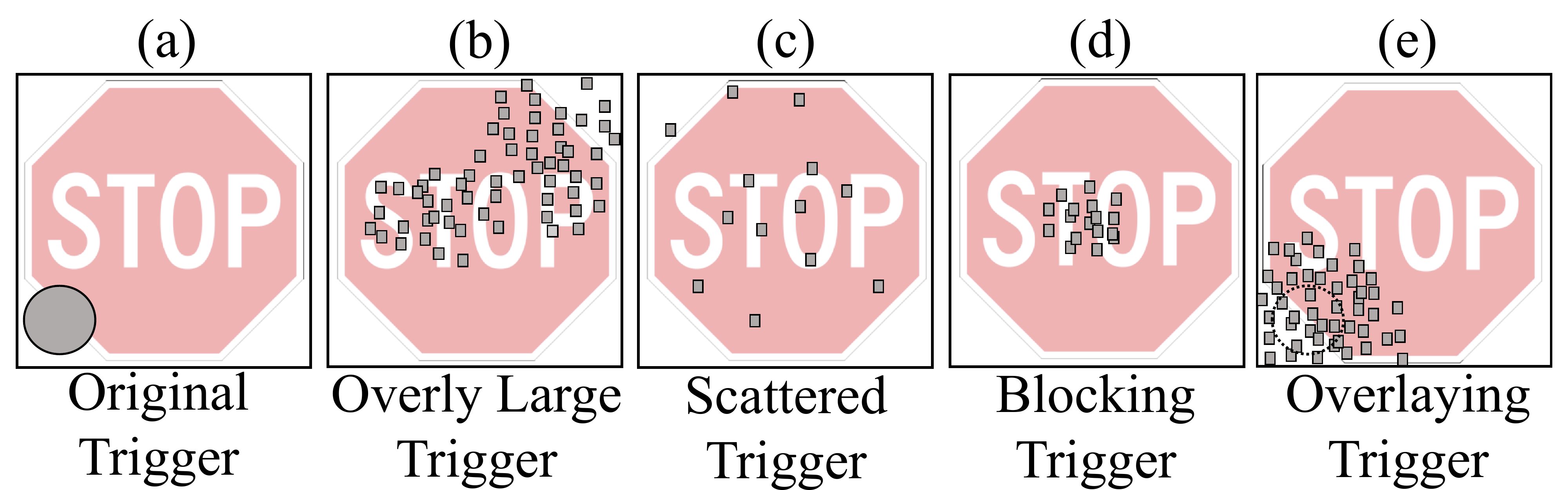}\\
  \end{center}
  \vspace{-5pt}
  \caption{The illustration of observed false alarms and incorrect triggers.}
  \label{fig:trojai07}
\vspace{-10pt}
\end{figure}

For a clean model $f(\cdot)$, the local optimum indicates a false alarm but not a trigger pertaining to an intentionally-inserted trojan backdoor. For an infected model $f(\cdot)$, the local optimum may represent the trigger intentionally inserted or an incorrect trigger (\ie an unintentionally-inserted trigger tied to a non-target class or a trigger tied to the target class but with nearly no overlaps with the trigger intentionally inserted). 

As is mentioned in Section~\S\ref{subsec:scope}, a model user aims to \ding{182} point out whether a model truly carries a trojan backdoor intentionally implanted (but not those naturally existing) and \ding{183} if an intentionally inserted trojan exists, restore the corresponding trigger as accurate as possible. To achieve the goals above, a trojan detection technique therefore has to minimize the negative influence of false alarms as well as incorrect triggers. In this work, we tackle this issue as follows. First, we conduct an empirical study to analyze the false alarms as well as incorrect triggers. Guided by our analysis, we then design a new optimization function (\ie objective function) to reduce the amount of unexpected local optima. Last, we introduce a new metric to evaluate the quality of the trigger restored and thus completely eliminate the influence of the remaining unexpected local optima upon trojan detection and trigger restoration. For the rest of this section, we first describe our observation. Then, we specify the design of our objective function as well as the metric. Finally, we present the strategy used for solving our optimization. 

\subsection{Observations}
\label{subsec:obs}


We trained one clean classifier and one infected classifier for the application of traffic sign recognition. Then, we detected trojan backdoors for both models by using the aforementioned optimization (\ie Equation~\eqref{eq:1}). Through this setup, we collected various triggers resolved. These include the trigger tied to the trojan intentionally inserted, the false alarms and incorrect triggers. In this work, we analyze these triggers and summarize their common characteristics below. As we will discuss in Section~\S\ref{subsec:details}, we design our objective function under the guidance of these characteristics.

\para{Observation I: Scattered \& Overly Large. } 
By observing false alarms and incorrect triggers, we first discover their presentations are either extremely scattered (see Figure~\ref{fig:trojai07} (c)) or overly large (see Figure~\ref{fig:trojai07} (b)). For those overly large, we also observe that they typically do not have overlaps with the trigger pertaining to the inserted trojan. By inserting any of incorrect triggers or false alarms into a set of clean images, we observe some of the trigger-inserted images could trick the corresponding model into yielding incorrect classification results. This implies that, the images tied to misclassification are adversarial samples, and both incorrect triggers and false alarms are triggers pertaining to trojan backdoors naturally existing. 

\para{Observation II: Blocking Key Object. } 
From the corpus of incorrect triggers and false alarms, we also find, for some of them, they are present with a decent density and in a reasonable size. As is depicted in Figure~\ref{fig:trojai07} (d), in terms of the size and shape, they look like a trigger intentionally inserted. However, inserting it into an image of a traffic sign, it inevitably blocks the key object in the image. Based on the statement in Section~\S\ref{subsec:scope} -- a trigger pertaining to an inserted backdoor has to be present at the corner of an image -- it cannot possibly be a trigger tied to a trojan backdoor intentionally inserted. Rather, it is merely a trigger pertaining to yet another trojan backdoor naturally existing. 

\para{Observation III: Overlaying. }
As is mentioned above, we also gathered the resolved trigger pertaining to the trojan intentionally inserted. By observing that resolved trigger, we surprisingly discover that it overlays the trigger intentionally inserted but presents itself in a larger size (see Figure~\ref{fig:trojai07} (e)). With respect to the task of detecting the existence of a trojan backdoor, this resolved trigger indicates the correct detection of a trojan backdoor. Regarding the task of restoring a trigger, however, this resolved trigger implies a restoration with a relatively low fidelity because it does not perfectly overlap the real trigger tied to the inserted trojan.  


\begin{figure*}[t]
  \begin{center}
  \includegraphics[width=0.95\textwidth]{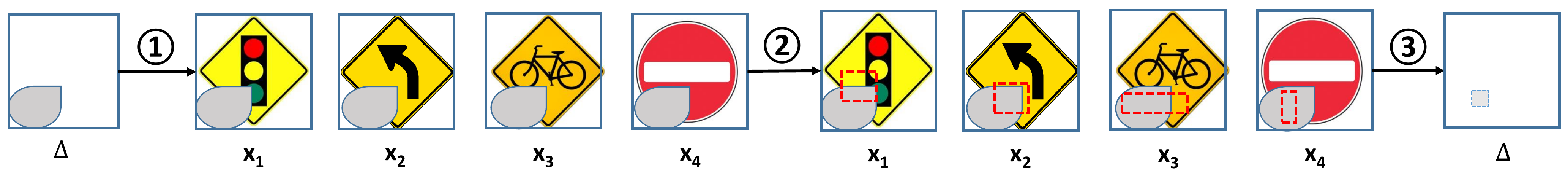}\\
  \end{center}
  \caption{The illustration of knocking off irrelevant features that are part of identified trojan backdoor. Note that the red box indicates the important features pinpointed through an explanation AI technique.}
  \label{fig:trojai04}
\vspace{-5pt}
\end{figure*}

\vspace{-5pt}
\subsection{Technical details}
\label{subsec:details}

To detect the existence of a trojan backdoor correctly and restore its corresponding trigger accurately, we propose \sys, an effective trojan detection approach. Technically, it first uses a new objective function to resolve triggers pertaining to intentionally-inserted trojan. Then, it leverages a new metric to analyze resolved triggers and thus determine the (non-)existence of a trojan backdoor. In this work, we design the new objective function by introducing four regularization terms under the guidance of our observations above, and design the new metric by using the heuristics established by those regularization terms. 
\vspace{-5pt}
\subsubsection{Regularization responding Observation I}

As is mentioned above, for false alarms and incorrect triggers, one of their characteristics is their overly-large sizes and scattered shapes. Therefore, when resolving the aforementioned optimization, we can introduce two regularization terms to penalize the resolved triggers overly large or scattered. Technically speaking, for penalizing the triggers overly large, it can be interpreted as controlling the size of the trigger resolved or, in other words, restricting the number of non-zero elements in $\bM$. For penalizing the triggers scattered, it can be viewed as restricting the smoothness of a trigger. With both of the penalization, when searching solutions in the adversarial subspace, we can ensure the optimization could have better chance to converge at a local optimum indicating the trigger pertaining to the inserted trojan. Conceptually, this is because, the constraints imposed to the optimization remove all the adversarial samples carrying an overly-large (or scattered) triggers and thus reduce the total number of adversarial samples in the adversarial subspace.

\para{Regularization term for overly large triggers. }
We define the regularization term pertaining to the penalization of overly large triggers as follows:
\begin{equation}
\label{eq:2}
\begin{aligned}
    R_{1}(\bM, \bd) &= \lambda_1 \cdot R_{\text{elastic}}(\text{vec}(\bM)) + \lambda_2 \cdot R_{\text{elastic}}(\text{vec}(\bd')) \, , \\
    \bd' &=  (\mathbf{1} - \bM) \odot \bd \, .
\end{aligned}
\end{equation}
Here, $\text{vec}(\cdot)$ denotes converting a matrix into a vector. $R_{\text{elastic}}(\cdot)$ represents imposing an elastic net~\cite{zou2005regularization} (the sum of $L_{1}$ and $L_{2}$ norms) to a vector. $\lambda_{1}$ and $\lambda_{2}$ are hyperparameters indicating the weights assigned to corresponding terms.

As we can observe in this equation, $\bM$ represents the mask of a trigger and $\bd'$ indicates the color pattern outside the region of that trigger. Imposing an elastic net on both of these terms and then summing them together, we can obtain a measure indicating the total amount of non-zero elements in $\bM$ and $\bd'$. For an overly large trigger, the value of $R_{1}(\bM, \bd)$ is relatively large. Therefore, by adding this regularization term into the Equation~\eqref{eq:1}, we can easily remove those triggers overly large and thus shrink the amount of adversarial samples in adversarial subspace. 


\para{Regularization term for scattered triggers. } 
We design the regularization term pertaining to the penalization of scattered triggers as follows:
\begin{equation}
\label{eq:3}
\begin{aligned}
    R_{2}(\bM, \bd) &= \lambda_3 \cdot s(\bM) + \lambda_4 \cdot s(\bd') \, , \\
    s(\bM) &= \sum_{i,j} (\bM_{i,j} - \bM_{i,j+1})^2 + \sum_{i,j} (\bM_{i,j} - \bM_{i+1,j})^2 \, .
\end{aligned}
\end{equation}
Here, $\lambda_3$ and $\lambda_4$ are hyperparameters indicating the weights to the corresponding terms. $\bM_{i,j}$ denotes the element in the mask matrix $\bM$ at the $i^{th}$ row and $j^{th}$ column. $s(\cdot)$ is a smoothness measure. Applying it to $\bM$ and $\bd'$, it describes the density of zero and non-zero elements. It is not difficult to imagine, the lower this smoothness measure is, the less scattered the resolved triggers will be. Together with the regularization term $R_{1}$, we add this new regularization term $R_{2}$ into the Equation~\eqref{eq:1}. With this term, we can eliminate the scattered triggers and further reduce the amount of adversarial samples in the adversarial subspace.

\subsubsection{Regularization responding Observation II}

Recall that we observed resolved triggers blocking the key object in images. To eliminate triggers in this category, we leverage the definition of trojan backdoors. As is mentioned in Section~\S\ref{subsec:scope}, in order to avoid blocking key objects in an input image, a malicious trigger is usually present at the corner of an input image. Therefore, by simply removing the trigger (or technically speaking nullifying the corresponding pixels) from a trigger-inserted image, a model should be able to correctly classify that trigger-removal image. In this work, we utilize this definition to derive a heuristic, and then design a regularization term by using this heuristic. We integrate this regularization into optimization and thus eliminate the images with blocking triggers from adversarial subspace. 

\para{Regularization term for blocking triggers. } 
Guided by the observation II, we design the regularization term pertaining to the blocking triggers as follows:
\begin{equation}
\label{eq:4}
\begin{aligned}
    R_3 = \lambda_5 \cdot L(f(\bx \odot (\mathbf{1} - \bM )), y_{t'}). \\
\end{aligned}
\end{equation}
Similar to the two regularization terms mentioned above, $\lambda_{5}$ is a hyperparameter controlling the weight assigned to the term. $\bx \odot(\mathbf{1} - \bM )$ represents an image from which we crop the corresponding trigger. $y_{t'}$ stands for the true class of $\bx$. As we can see from the equation above, this regularization term  $R_{3}$ introduces a loss function $L(\cdot)$ to measure the similarity between the true class and the prediction for $\bx \odot(\mathbf{1} - \bM )$. By adding the term into the optimization shown in the Equation~\eqref{eq:1}, we can minimize the similarity and thus eliminate those blocking triggers effectively. Conceptually, this implies removing images with blocking triggers from adversarial subspace.

\subsubsection{Regularization responding Observation III}

In addition to the blocking triggers, recall that we also observe triggers that overlay the trigger pertaining to the trojan intentionally inserted. As is discussed in Section~\S\ref{subsec:obs} (Observation III), while this kind of triggers is correctly tied to the trigger of our interest, they do not represent a high-fidelity restoration. To address this problem and improve fidelity for resolved triggers, we design a regularization term by borrowing the idea of those explainable AI techniques (\eg~\cite{fong2017interpretable, dabkowski2017real, guo-2018-neurips, guo-2018-ccs, simonyan2013deep, singla2019understanding, sundararajan2017axiomatic, ribeiro2016should, melis-2018-neurips}). The rationale behind our design is as follows. 

An overlaying trigger indicates a trigger with additional irrelevant features enclosed. For an input with an overlaying trigger, the misclassification is dominated by those features tied to the actual trigger pertaining to the inserted trojan. Therefore, the most effective approach to addressing overlaying triggers is to knock off all the irrelevant features, preserve the minimal set of features attributive to the prediction result and deem the remaining set as the trigger intentionally inserted. For explainable AI techniques, their goal is to identify the most important features contributing to a classification result. Given an input sample $\bx_{t}$ with a resolved trigger attached, an explainable AI technique therefore could assess the feature importance, eliminate those features not dominant to the misclassification and eventually provide us with a trigger in a higher fidelity. 


\para{Regularization term for overlaying triggers. } 
Given a prediction result $y$ for a particular input $\mathbf{x}$, an explainable AI technique (\eg~\cite{fong2017interpretable, dabkowski2017real}) could pinpoint the top important features that contribute to the prediction result by solving the following function:
\begin{equation}
\label{eq:5}
\begin{aligned}
    \text{argmin}_{\bM_1} L(f(\bx \odot\bM_1 ), y) \, .\\
\end{aligned}
\end{equation}
Here, $M_1$ is an explanation matrix with the same dimensionality as the input $x$. Each of its elements is either $0$ or $1$. Using the function above, one could find a minimal set of features for $\bx$ that contributes most to the prediction result $y$. For many explainable AI research works, the `one' elements in $\bM_1$ depict the minimal set of features, and $\bx \odot\bM_1$ indicates the explanation for the prediction $y$. 

According to the definition of trojan backdoors, a trigger $(\bM \odot \bd)$ present in an arbitrary clean input $\mathbf{x}$ could mislead the infected classifier into categorizing the input data from its original label to a target label $y_{t}$. This implies that the trigger $(\bM \odot \bd)$ should be the most important features for all the input samples with the trigger attached. Based on this property, intuition suggests that after solving the objective function~\eqref{eq:1}, we can follow the steps below to knock off irrelevant features in a restored trigger. 

First, we add the restored trigger $({\bM} \odot {\bd})$ back to the testing samples and obtain a set of bad samples (see {\ding{192}} in Figure~\ref{fig:trojai04}). Second, we utilize an explanation approach to pinpoint the top important features for each of the bad samples (see {\ding{193}} in Figure~\ref{fig:trojai04}). Since for each bad sample, we identify a set of important features, third, we employ a majority vote mechanism to select the features that mostly appear in all sets of important features (see {\ding{194}} in Figure~\ref{fig:trojai04}). Finally, we deem the features selected as the trojan backdoor originally inserted.

By following the steps above, we can obtain a trigger in a higher fidelity. In this work, we however model the aforementioned procedure as a regularization term below and integrate it into the optimization above because this could ease our computation. 
\begin{equation}
\label{eq:6}
\begin{aligned}
    R_{4} = & \lambda_{6}\cdot L(f(\bx_t \odot\bM_1 ), y_t)  \\
     = & \lambda_{6}\cdot L(f((\bx \odot (\mathbf{1} - \bM) + \bM \odot \bd) \odot\bM_1 ), y_t)  \\
     = & \lambda_{6}\cdot L(f((\bx \odot (\mathbf{1} - \bM)) \odot \bM_1 + (\bM \odot \bd) \odot\bM_1), y_t)\, , \\
\end{aligned}
\end{equation}
As we can observe, the regularization term above is a loss function. By plugging it into the Equation~\eqref{eq:1} together with the other three regularization terms mentioned above, we introduce Equation~\eqref{eq:5} into the objective function used for resolving malicious triggers. By setting $\bM = \bM_{1}$, we can force the optimization to compute a trigger $\bM$ which is the most important features for $\bx_{t}$. It should be noted that, the elements of both $\bM$ and $\bM_1$ are designed to be either $1$ or $0$ and, therefore, we can easily derive $(\bx \odot (\mathbf{1} - \bM)) \odot \bM_1 = \mathbf{0}$ and $(\bM \odot \bd) \odot\bM_1 = \bM \odot \bd$, and eventually obtain a simplified form for the regularization term $R_{4}$, \ie
\begin{equation}
\label{eq:7}
\begin{aligned}
R_4 = \lambda_6 \cdot L(f(\bM \odot \bd), y_t)\, , \\
\end{aligned}
\end{equation}
As is mentioned above, together with the other three regularization terms, we include this regularization as part of the optimization depicted in Equation~\eqref{eq:1}. In Appendix, we specify the complete optimization function which integrates all the regularization terms.

\subsubsection{Metric Design}



Using the optimization function (shown in the Equation~\eqref{eq:1}) augmented with the four regularization terms mentioned above, we can significantly reduce the amount of adversarial samples in the adversarial subspace and potentially increase the possibility of obtaining the trigger truly tied to the trojan inserted. However, similar to the optimization problem alone defined in the Equation~\eqref{eq:1}, for any given value $y_{t}$ and a deep neural network model $f(\cdot)$, we still inevitably obtain a solution (a local optimum). In the pioneering work~\cite{wang-2019-ieeesp}, researchers have computed the $L_{1}$ norm for each local optimum and then utilized an outlier detection method (MAD~\cite{leys2013detecting}) to filter out local optima indicating incorrect triggers and false alarms. However, as we will show in Section~\S\ref{sec:eval}, $L_{1}$ norm provides the outlier algorithm with extremely poor distinguishability. In this work, we therefore use the MAD outlier detection method to eliminate false alarms and incorrect triggers but replace the $L_{1}$ norm with a new metric. In Section~\ref{sec:eval}, we will show, with this new metric and our regularization, we significantly improve the performance for trojan detection and trigger restoration.

Mathematically, our new metric is defined as follow:
\begin{equation}
\label{eq:11}
\begin{aligned}
    A(\bM_t, \bd_t) =  & \text{log}(\frac{\|\text{vec}(\bF^{(t)})\|_1}{d^2}) + \text{log}(\frac{s(\bF^{(t)})}{d\cdot(d-1)}) \\
                       &- \text{log}(\text{acc}_{\text{att}})  - \text{log}(\text{acc}_{\text{crop}}) - \text{log}(\text{acc}_{\text{exp}}) \, .
\end{aligned}
\end{equation}
For each trigger resolved (i.e., $\bM_t \odot \bd_t$), here, $\text{acc}_{\text{att}}$ indicates the misclassification rate, observed when we insert that resolved trigger into a set of clean images and then feed these contaminated images into the learning model $f(\cdot)$. $\text{acc}_{\text{exp}}$ represents the classification accuracy, observed when we feed the explanation (\ie important features alone) of the contaminated images into $f(\cdot)$. $\text{acc}_{\text{crop}}$ denotes the classification accuracy, observed when we crop resolved triggers from the corresponding contaminated images and then input the cropped images to $f(\cdot)$. For the first two terms in the equation above, we define $\bF^{(t)}_{ij} = \mathbf{1}\{(\bM_t \odot \bd_t)_{ij} > 0 \}$. With respect to $\|\text{vec}(\bF^{(t)})\|_1$ and $s(\bF^{(t)})$, they represent the sparsity measure indicating non-zero elements in the trigger and the smoothness measure of the trigger, respectively. As is mentioned in the beginning of this section, we assume an input $\bx$ is in a $d \times d$ dimension (\ie $\bx \in \mathbb{R}^{d \times d}$). In the equation above, we use $d^2$ and $d\cdot(d-1)$ to normalize both the sparsity and smoothness measures into a range of $0 \sim 1$. 


\subsection{Strategies for Resolving Optimization}
Here, we discuss the strategies used for resolving the aforementioned optimization.

\para{Optimization algorithm. }
Given a set of testing samples and their corresponding labels, we resolve our optimization function by using a gradient decent approach. To be specific, we adopt {\tt Adam}~\cite{kingma2014adam} to accelerate the optimization process needed for gradient descent computation. Different from the Vanilla gradient decent, which has a fixed learning rate, {\tt Adam} adjusts the learning rate via dividing it by an exponentially decaying average of squared gradients. More details related to this optimization approach can be found in~\cite{kingma2014adam}. 

\para{Hyperparameter augmentation. }
As is described above, we introduce $6$ hyperparameters (\ie $\lambda_1 \sim \lambda_6$). To reduce the influence of these hyperparameters upon the trojan detection performance, we introduce a hyperparameter augmentation mechanism as follows. First, we initialize each $\lambda_i$ with a relatively small value. This indicates that, at an early stage, we do not heavily use our regularization terms for trigger restoration. Second, we resolve optimization and then insert the trigger (\ie local optimum) into a set of clean input samples. Third, we feed the trigger-inserted images into the corresponding learning model and then measure the misclassification rate. If the misclassification rate reaches a certain threshold $\phi$, we increase each of the hyperparameters by multiplying it with a step variable $m$. Otherwise, we divide the current $\lambda_i$ by that step variable $m$. In this work, we iterate this procedure until each of the aforementioned regularization terms stays stable (\ie $|R_t^{(k-1)} - R_t^{(k)}| < \epsilon$, where $R_t^{(k)}$ is the value of the $t^{th}$ regularization term at the $k^{th}$ iteration). As we will demonstrate in Section~\S~\ref{sec:eval}, our hyperparameter augmentation mechanism enables a stable detection result when each hyperparameter is initialized within a certain range.   


\begin{figure*}[t]
    \centering
    \includegraphics[width=0.99\textwidth]{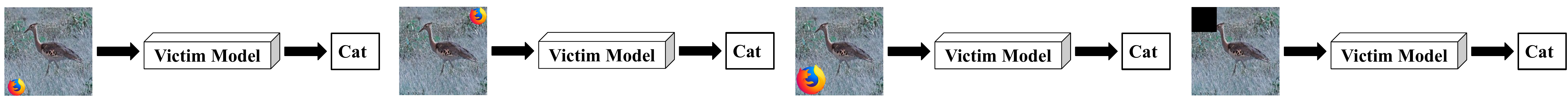}
    \vspace{-5pt}
    \caption{The demonstrations of the victim models that are trained on ImageNet dataset and are infected by the triggers with different shapes, locations and sizes.} \label{fig:diff_setting}
    \vspace{-5pt}
\end{figure*}

\commenta{
\begin{figure}[t]
  \begin{center}
  \includegraphics[width=0.45\textwidth]{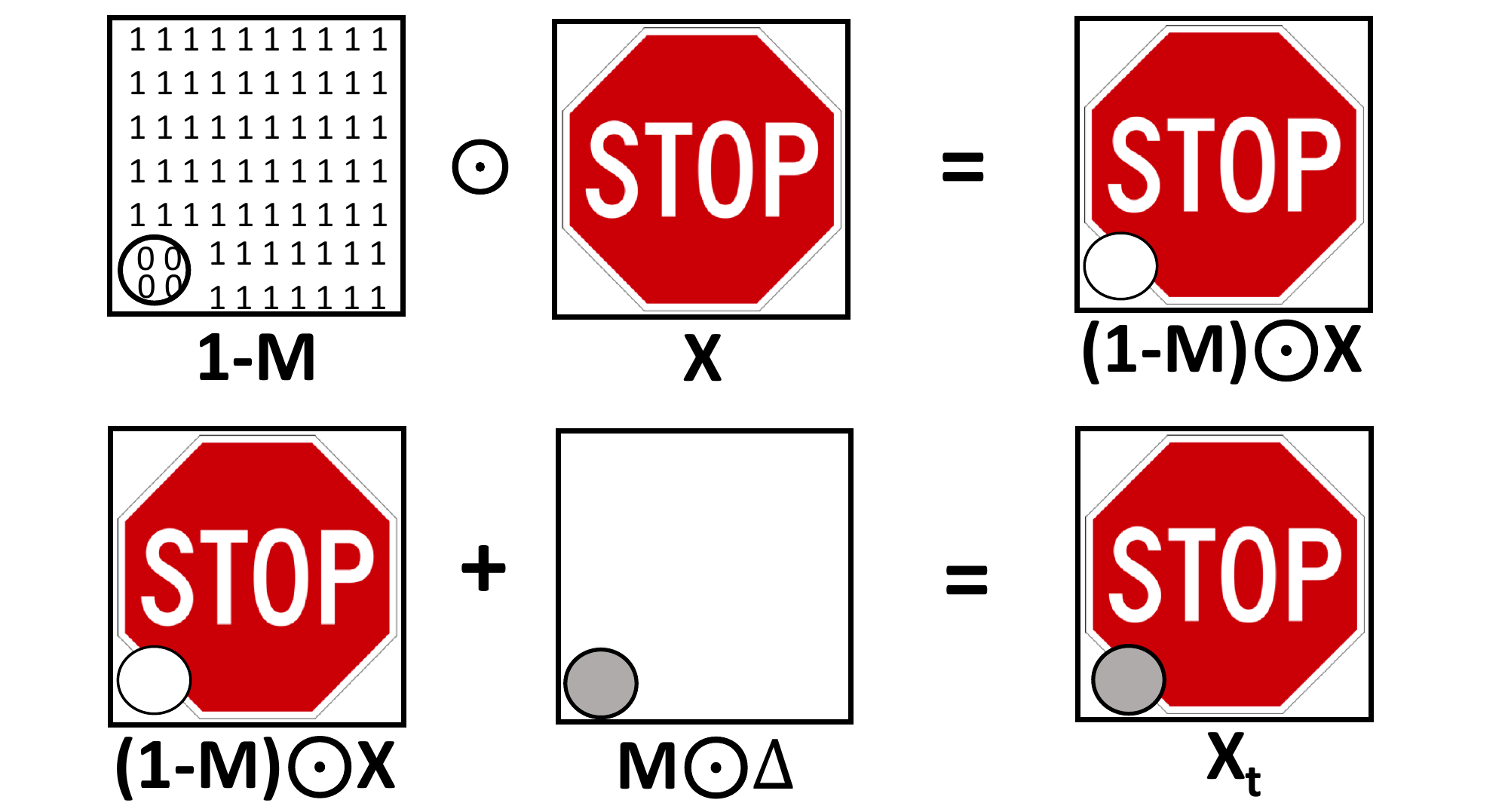}\\
  \end{center}
  \caption{The illustration of trigger insertion. Note that the yellow mark is the trigger and $\bM$ is a mask matrix. The matrix has the dimensionality of $d \times d$ with the elements in the trigger-presented-region equal to `1' whereas all the others equal to `0'.}
  \label{fig:trojai01}
\end{figure}

\section{Key Technique}
\label{sec:tech}

To address the limitations of existing techniques, we propose \sys to detecting trojan backdoors intentionally inserted into a victim learning model. At a high level, \sys restores a trigger (\ie $\bM$ and $\bd$) from every possible output class of an infected model via solving an optimization problem and then applies an outlier detection method to identify the real triggers in contaminated classes and false alarms (\ie triggers restored from clean classes). Technically speaking, given an infected model, restoring an injected trigger can be viewed as computing a mask $\bM$ and a pattern $\bd$ from that model, where $\bM$ (indicating the shape and the location of the injected trigger) and $\bd$ (representing the color pattern of the trigger) together decide the restored trigger (see $\bM\odot\bd$ in Figure~\ref{fig:trojai01}). To be specific, $\bM$ and $\bd$ can be resolved from the following optimization function:
\begin{equation}
\label{eq:1}
\begin{aligned}
    \text{argmin}_{\bd, \bM} L(f(\bx_t), y_{t}) \, , \\
    \bx_t = \bx \odot (\mathbf{1} - \bM) + \bd \odot \bM \, , 
\end{aligned}
\end{equation}
where $\bx \in \mathbb{R}^{d \times d}$ is a testing sample in matrix form\footnote{Note that if an input is a colored image, $\bx$ should be a 3-D tensor of ${R}^{d \times d \times 3}$ instead of a 2-D matrix.} and $\bx_t$ denotes the data sample $\bx$ with the trigger $\bd \odot \bM$ inserted at the area indicated by mask $\bM$ (An example is shown in Figure~\ref{fig:trojai01}). $f(\cdot)$ represents the infected model. The loss function $L(f(\bx_t), y_{t})$ indicates the similarity between the prediction $f(\bx_t)$ and the target label $y_{t}$. Here, $y_{t}$ represents a specific target label which an infected model misclassifies arbitrary contaminated inputs (\ie $\bx_t$) into. 

Solving Equation~\eqref{eq:1} is equivalent to finding a local optimum from the entire adversarial space of $f(\cdot)$ and deem that optimum as the trojan backdoor intentionally inserted. However, due to non-convex or non-concave proprieties of DNN~\cite{goodfellow2016deep}, it usually has multiple local optima for an objective function. As a result, the solutions of Equation~\eqref{eq:1} frequently fall into other local optima (\ie adversarial samples) instead of the desired one (\ie trojan backdoor). \fixme{Actually, we have observed four failure cases when solving Equation~\eqref{eq:1}. As we will show later in Section\S~\ref{sec:eval}, these failure cases will then lead to poor end-to-end detection performance and low fidelity of restored triggers. To avoid these cases and improve the trojan detection accuracy, we design a new optimization function by introducing four kinds of regularization terms to Equation~\eqref{eq:1}. In the following, we first describe our observations (\ie failure cases) when solving Equation~\eqref{eq:1}, followed by our technical details and the strategies we adopt during solving the optimization function.}
\begin{figure}[t]
  \begin{center}
  \includegraphics[width=0.45\textwidth]{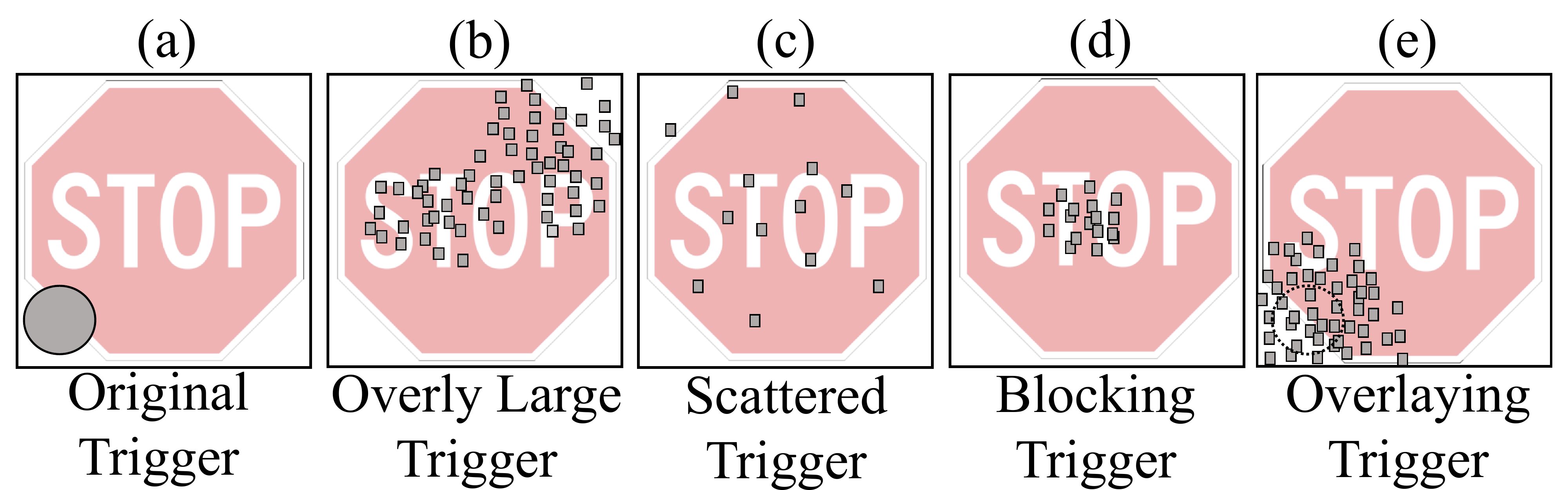}\\
  \end{center}
  \caption{The illustration of observed failure cases. \fixme{I suggest also add the stop sign to the remaining four figures. remember to fix the names in the figure.}}
  \label{fig:trojai07}
\end{figure}
\subsection{Observations}

\fixme{the name of the four categories.}
In order to mitigate the problem of adversarial samples (adv for short), we first conducted an empirical study on the triggers or adversarial samples restored from contaminated classes or clean classes using Equation~\eqref{eq:1}. Intuitively speaking, an adversarial sample should achieve two goals: (1) cover the important. We observed that failure cases in trigger restoration can be roughly divided into four categories: \textcolor{red}{\ding{192} overly large adv, \ding{193} overly large trigger, \ding{194} overlapping adv, and \ding{195} scattered adv. \ding{192} and \ding{193} usually happen in a contaminated class, and \ding{194} and \ding{195} occur in clean classes. In the rest of this subsection, we describe each case in details and discuss the reason for them respectively.}


Recall that a contaminated class in an infected model may have multiple local optima for Equation~\eqref{eq:1}, \textcolor{red}{some of which lie in a relatively higher dimensional subspace than the others including the one representing the injected backdoor.} If Equation~\eqref{eq:1} falls into such a local optimum, it will return a \emph{overly large trigger} (\ie a trigger that is larger than the real trojan). Due to the non-linearity of DNN, an overly large trigger is usually randomly distributed and it frequently has a small overlap with the real injected trigger (See Figure~\ref{fig:trojai07} for an example). Because overly large trigger is similar to a false alarm, it will cause a detection approach to falsely treat an infected class as a clean class. Despite the optimization process avoids overly large triggers, it is still extremely hard to restore a trigger with almost the same size as the original one. Usually, what we get is a \emph{covered trigger} that cover the region of the true trojan but including some false positive features (shown in Figure~\ref{fig:trojai07}). This is because that a trigger that slightly larger than the original injected trigger can also cause the infected model to behave misclassification as long as they share great overlapping regions. As a result, the trigger that restores from~\eqref{eq:1} sometime is a super-set of the trigger originally implanted. Similar to overly large trigger, covered trigger will cause the trigger restored from the infected label to be less distinguishable with the false alarms and make the detection approach fail to pinpoint the contaminated classes.

Besides the former two cases that occur in an infected class, another two cases that commonly happens in a clean class are overlapping trigger and scatter trigger (shown in Figure~\ref{fig:trojai07}). \emph{Overlapping trigger} stands for a trigger that has an overlapping part with the object in an input image and \emph{Scatter trigger} refers to the case that the restored trigger is scattered and sparse. These two false alarms usually include a subset of features in the object of an input image. This is because the classification result of input is highly related to features in the object, which indicates that they are more sensitive than other features in that a slight manipulation of them will lead the model to misclassify that input sample. In these cases, solving Equation~\eqref{eq:1} is similar to generating an adversarial sample. As is introduced in~\cite{carlini2017towards}, an adversarial sample can be generated by perturbing a small portion of the sensitive features (\ie features in the object) or a few very important ones, which corresponds to overlapping trigger and scatter trigger.  As is demonstrated in Figure~\ref{fig:trojai07}, an overlapping trigger, and a scatter trigger are similar to an original trigger in that all of them occupy a small proportion of an input image. As a result, these two false alarms will highly probably cause the detection approach to returning a false positive class.

\subsection{Technical details}
To filter out these false restored triggers, we design four different regularization terms for Equation~\eqref{eq:1}. In the following, we introduce the technical details of the designed terms and the anomaly detection mechanism we adopt to identify the contaminated classes. 

\para{Avoiding overly large triggers.} 
To avoid the adversarial space where overly large trigger rides in, we need to constrain the size of a restored trigger (\ie the number of non-zero elements in $\bM$). In other words, we want the restored trigger mask $\bM$ to be sparse. Another reason for ensuring a sparse mask is that by definition, a trojan backdoor only occupies a relatively small region within an input. To achieve this goal, former approach~\cite{wang-2019-ieeesp} add an $L_{1}$ regularization term on the $\text{vec}(\bM)$\footnote{\text{vec(.)} transforms matrix $\bM$ into a vector~\cite{macedo2013typing}, because the definition of $\|\cdot\|_1$ and  $\|\cdot\|_2$ in a matrix space is different from those defined in a vector space.}, which constrains the sum of each element in a $\bM$. However, for a high dimensional input (\eg colored image of size $32\times32$), $L_{1}$ actually sacrifices the sparsity restriction because a low value of $L_{1}$ could not only represent a highly sparse but also a low-sparsity matrix with most elements equal to a tiny non-zero value. As a result, instead of $L_{1}$ norm, we choose to use elastic-net $\|\text{vec}(\bM)\|_{1} + \|\text{vec}(\bM)\|_{2}$ to achieve the sparsity. Elastic-net is the sum of the $L_{1}$ norm and the $L_{2}$ norm of $\text{vec}(\bM)$. In high dimensional space, elastic-net could achieve higher sparsity than $L_{1}$ norm~\cite{zou2005regularization}. 

Beside forcing $\bM$ to be sparse, we can further restrict the size of the restored trojan $\bM\odot\bd$ by adding sparsity constraint on $(\mathbf{1} - \bM) \odot \bd$. Recall that an injected trojan only has a certain color pattern occupying a small part of the whole image and the remain parts (\ie $(\mathbf{1} - \bM) \odot \bd$) is background, which can be treated as zero values (See Figure~\ref{fig:trojai01} for an example). This indicates that  remain part of the recovered trigger should be very sparse. As a result, we can also introduce elastic-net constraint on $(\mathbf{1} - \bM) \odot \bd$. To be specific, we can introduce a sparsity regularization $R_s$ with the following form:
\begin{equation}
\label{eq:2}
\begin{aligned}
    R_{1}(\bM, \bd) &= \lambda_1 R_{\text{sparse}}(\bM) + \lambda_2 R_{\text{sparse}}(\bd') \, , \\
    \bd' &=  (\mathbf{1} - \bM) \odot \bd \, .
\end{aligned}
\end{equation}
where $R_{\text{sparse}}(\cdot)$ indicates the elastic-net.

\begin{figure*}[t]
\hfill
\begin{minipage}{.65\textwidth}
  \centering
  \includegraphics[width=0.9\textwidth]{img_tech/trojai04.pdf}\\
  \caption{The illustration of knocking off irrelevant features that are part of identified trojan backdoor. Note that the red box indicates the important features pinpointed through an explanation AI technique.}
    \label{fig:trojai04}
\end{minipage}
\hspace{20pt}
\begin{minipage}{0.3\textwidth}
  \centering
  \includegraphics[width=0.9\textwidth]{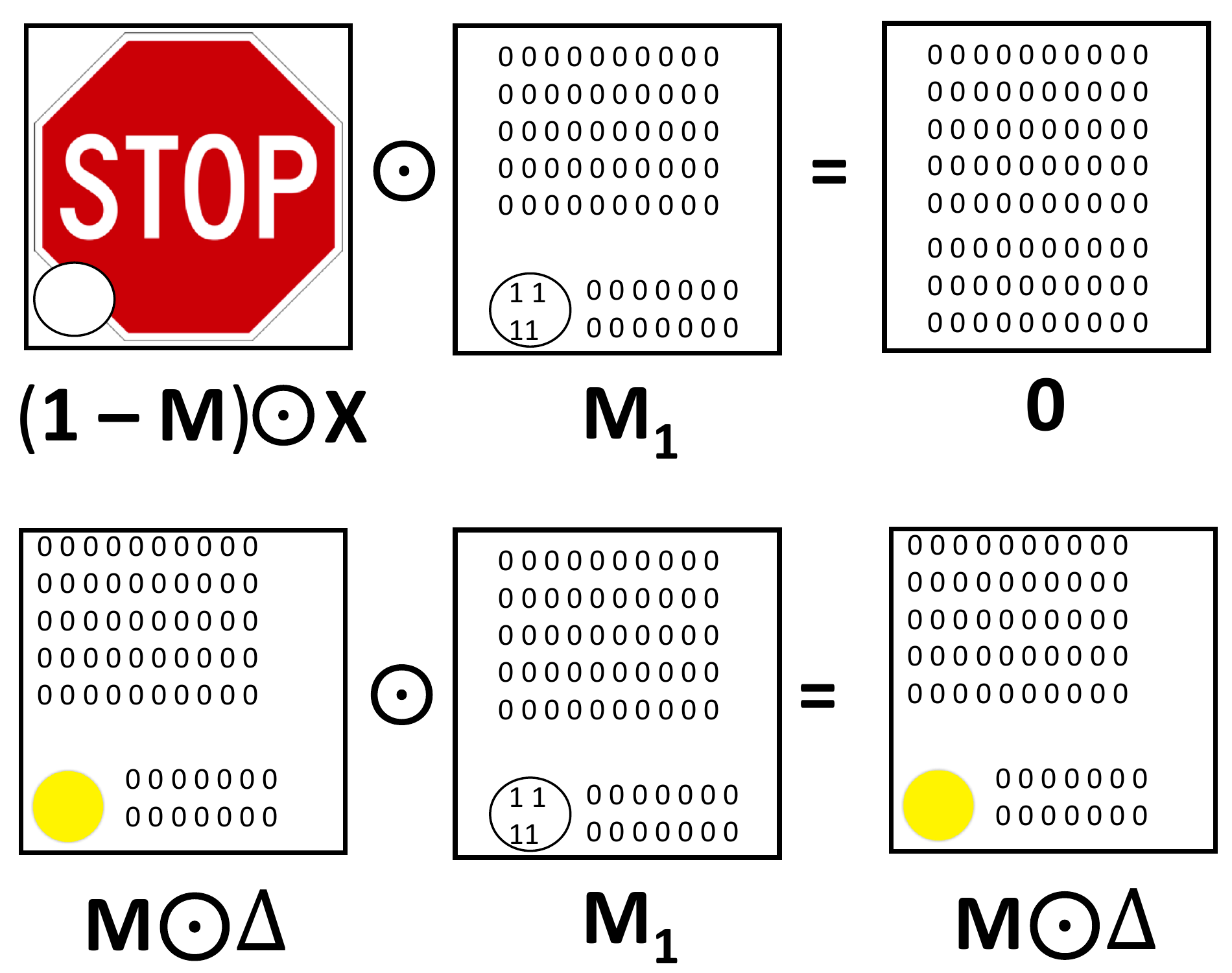}\\
  \caption{The demonstration of Equation~\eqref{eq:5}. Here $\bM = \bM_1$.}
  \label{fig:trojai06}
\end{minipage}
\end{figure*}

\para{Trimming covered triggers.} 
To filtering our the false discovery features in covered triggers, we seek help from explainable AI techniques~\cite{fong2017interpretable,dabkowski2017real,guo-2018-neurips,guo-2018-ccs,simonyan2013deep,singla2019understanding,sundararajan2017axiomatic,ribeiro2016should,melis-2018-neurips} to knock off the irrelevant features that are part of the triggers identified. Given a prediction result $y$ for a particular input $\mathbf{x}$, an explainable AI technique (\eg~\cite{fong2017interpretable, dabkowski2017real}) could pinpoint the top important features that contribute to the prediction result by solving the following optimization function:
\begin{equation}
\label{eq:3}
\begin{aligned}
    \text{argmin}_{\bM_1} L(f(\bx \odot\bM_1 ), y) + \lambda ( \|\text{vec}(\bM_1)\|_{1} + \|\text{vec}(\bM_1)\|_{2})\, .\\
\end{aligned}
\end{equation}
This optimization function intend to find a minimum part of the original input sample $\bx$ (represented by $\bM_1$) that contributes most to the prediction result. Then the masked sample ($\bx \odot\bM_1 $) can be treated as the explanation of $\bx$. According to the definition of trojan backdoors, a trigger $(\bM \odot \bd)$ presented in an arbitrary clean input $\mathbf{x}$ could mislead the infected classifier into categorizing the input data from its original label to a target label $y_{t}$. This implies that the trigger $(\bM \odot \bd)$ should be the most important features for all the input samples with the trigger presented. 

\textcolor{red}{replace ding with a better looking circled number.}
Based on this property, intuition suggest that after solving the objective function~\eqref{eq:1}, we can first add the restored trigger $(\hat{\bM} \odot \hat{\bd})$ back to the testing samples and obtain a set of infected samples (see {\ding{192}} in Figure~\ref{fig:trojai04}).  we can then utilize an explanation approach to pinpoint the top important features for each sample that cause that sample to be classified in the infected label $t$ (see {\ding{193}} in Figure~\ref{fig:trojai04}). Therefore, we can then employ a majority vote mechanism to select the features that mostly appear in all sets of important features (see {\ding{194}} in Figure~\ref{fig:trojai04} and deem the features selected as the trojan backdoor originally inserted. 

However, instead of performing this post-detection processing, we could integrating the explanation part into our detection objections. In other words, we build Equation~\eqref{eq:3} into the detection objection function~\eqref{eq:1}, which enable our approach to restore the implanted trigger and at the same time knockoff the false positive features via explanation. To be specific, according to Equation~\eqref{eq:3}, explaining an infected sample $\bx_t$ can be achieved by minimizing the following optimization function regularized by $\|\text{vec}(\bM_1)\|_{1} + \|\text{vec}(\bM_2)\|_{2}$ with respect to $\bM_1$:
\begin{equation}
\label{eq:4}
\begin{aligned}
    & L(f(\bx_t \odot\bM_1 ), y_t)  \\
     = & L(f((\bx \odot (\mathbf{1} - \bM) + \bM \odot \bd) \odot\bM_1 ), y_t)  \\
     = & L(f((\bx \odot (\mathbf{1} - \bM)) \odot \bM_1 + (\bM \odot \bd) \odot\bM_1), y_t)\, , \\
\end{aligned}
\end{equation}
where $\bM$ is the trigger mask and $\bM_1$ is the explanation mask. Ideally, we want $\bM = \bM_1$, which means the restored triggers are the most important features lead $\bx_t$ to be classified as $y_t$. Since the elements of both $\bM$ and $\bM_1$ are designed to be either $1$ or $0$, given $\bM = \bM_1$, we can derive that $(\bx \odot (\mathbf{1} - \bM)) \odot \bM_1 = \mathbf{0}$ and $(\bM \odot \bd) \odot\bM_1 = \bM \odot \bd$. Then, Equation~\eqref{eq:4} can be simplified as:
\begin{equation}
\label{eq:5}
\begin{aligned}
L(f(\bx_t \odot\bM_1 ), y_t) = L(f(\bM \odot \bd, y_t)\, , \\
\end{aligned}
\end{equation}
where $\bM = \bM_1$ (See Figure~\ref{fig:trojai06} for a demonstration). We then treat Equation~\eqref{eq:5} as the second regularization (\ie $R_2 = \lambda_3 L(f(\bM \odot \bd, y_t)$) and add it to Equation~\eqref{eq:1}. 

\para{Knocking off overlapping triggers.} 
As is shown in Fig~\ref{fig:trojai07}, an overlapping trigger usually has a large overlap with the object in an input. And the features in this region are of the important features that lead the model to classify that input into the right class. This indicates that if deleting the overlapping trigger from an input sample originally being correctly classified, the model will misclassify the remain parts. Given this property, we can add the following regularization into optimization function~\eqref{eq:1}: 
\begin{equation}
\label{eq:6}
\begin{aligned}
    R_3 = \lambda_4 L(f(\bx_ \odot\ (\mathbf{1} - \bM ), y_{t'}), \\
\end{aligned}
\end{equation}
where $\bx \odot\ (\mathbf{1} - \bM )$ represent the remain parts and $y_{t'}$ refer to the original class of $\bx$. 

\para{Eliminating scatter triggers.} 
Last but not least, to eliminate scatter trigger, we introduce a smoothness regularization on $s(\bM)$:
\begin{equation}
\label{eq:7}
\begin{aligned}
    R_{\text{smooth}}(\bM) = \sum_{i,j} (\bM_{i,j} - \bM_{i,j+1})^2 + \sum_{i,j} (\bM_{i,j} - \bM_{i+1,j})^2 \, ,
\end{aligned}
\end{equation}
indicating the overall density of the trigger. The more scattered trigger pixels are the higher the value the $s(\bM)$ has. For detecting a trojan backdoor implanted in an image classifier, the trigger in the infected class is typically presented in a dense and smooth fashion. However, the adversarial samples in a clean class (\ie scatter triggers) usually are more scatter especially for a complex model like DNN. For this task, we can, therefore, utilize this regularization to eliminate the influence of adversarial samples upon trojan detection. Similar to $R_1$, we can also further ensure the smoothness of the restored trigger $\bM \odot \bd$ by adding $R_{\text{smooth}}$ to $(\mathbf{1} - \bM) \odot \bd$. As is discussed before, this features in this part are all zeros and it must be highly smooth. We can now write the fourth regularization in the following form:
\begin{equation}
\label{eq:8}
\begin{aligned}
    R_{4}(\bM, \bd) = \lambda_5 R_{\text{smooth}}(\bM) + \lambda_6 R_{\text{smooth}}(\bd') \, , \\
\end{aligned}
\end{equation}
where $R_{\text{smooth}}(\cdot)$ refers to the smoothness regularization in Equation~\eqref{eq:7}.

}
\section{Experiment}
\label{sec:eval}

In this section, we evaluate the effectiveness of our proposed technique by answering the following questions. (1) Does the effectiveness of \sys vary when the size, position, and shape of trojan backdoors change? (2) Does the dimensionality of the input space tied to a target model influence the effectiveness of \sys? (3) How effectively does \sys demonstrate when there are multiple triggers inserted into target models? (4) Does the complexity of the target model influence the effect of trojan backdoor detection? (5) How sensitive is \sys to the hyperparameters? (6) How effective is \sys for different methods of inserting a trojan backdoor? (7) In comparison with the state-of-the-art technique -- {\tt Neural Cleanse}~\cite{wang-2019-ieeesp}, how well does \sys perform trojan  detection and trigger restoration under the settings above? In the following, we first describe the design and setup of our experiments. Then, we specify the datasets involved in our experiments as well as the evaluation metrics we use for our experiments. Finally, we show and discuss our experiment results.

\subsection{Experiment Design and Setup}

To answer the questions above, we design a series of experiments, each of which answers a subset of these questions. 

To answer the question (1), (2) and (7), we first selected two datasets GTSRB and ImageNet, on which we trained many neural networks with the architectures of 6 Conv+2 MaxPooling CNN for GTSRB and VGG16~\cite{simonyan2014very} for ImageNet (see Appendix for more details as well as other hyperparameters). In this way, we could obtain two sets of learning models, each of which takes inputs in a unique dimensionality ($32 \times 32 \times 3$ for GTSRB and $224 \times 224 \times 3$ for ImageNet). With these models in hand, we then contaminated these networks by using the trojan insertion approach BadNet~\cite{gu2017badnets}. 

BadNet implants a trojan backdoor into a target model, and the trigger tied to the inserted trojan is active only if a small, fixed-size trigger is present at the exact position of an image. With such a trojan backdoor, the target model misclassifies a trigger-attached image into the target class No.33 for models pertaining GTSRB and the target class No.1 for models pertaining to ImageNet. In this experiment, we adjust the size, shape and position of a trigger when using BadNet to train victim models\footnote{The details about the shape, location, and size of each implanted trigger are shown in Table~\ref{tab:pri}. It should be noted that we used square trojans with different colors (\ie white for GTSRB and black for ImageNet) for different datasets to test the robustness of the detection approaches to trojan colors.}. Figure~\ref{fig:diff_setting} showcases some learning models that misclassify an image only when the corresponding trigger is present in that image. In this work, we utilized \sys and {\tt Neural Cleanse} to detect trojan backdoors in victim models. In addition, we used both approaches to examine backdoor existence against clean models that have not been implanted with any trojan backdoors. 

\begin{figure}[ht]
    \centering
    \includegraphics[width=0.48\textwidth]{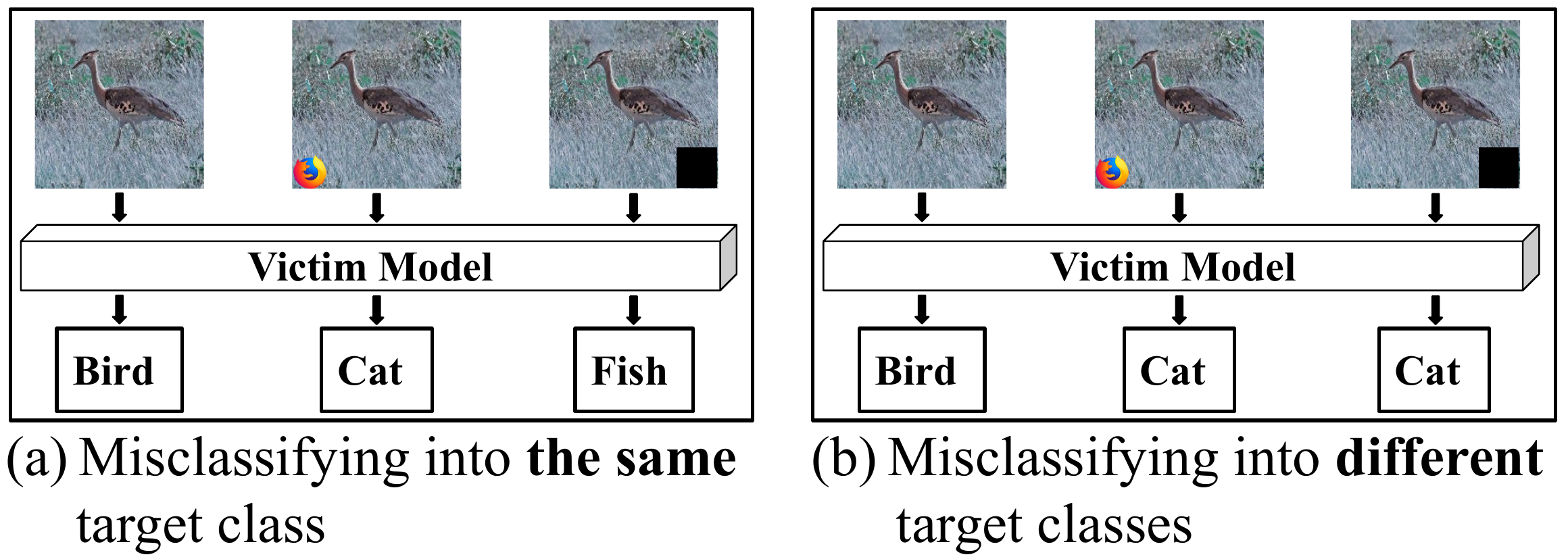}
    \vspace{-10pt}
    \caption{An example illustrating the effect of the two trojan backdoors upon the classification results.}
    \label{fig:2_demo}
    \vspace{-5pt}
\end{figure}

To answer the question (3) and (7), we trained the same neural networks on the aforementioned datasets and used BadNet to contaminate these networks. Differently, we, however, followed two alternative approaches to insert trojan backdoors into target models. For the first approach, we implant two backdoors into a single neural network. One backdoor is used for misclassifying an image (with the corresponding trigger A present) into a target class A. The other is used for mistakenly categorizing that image (with a different trigger B present) into another target class B. For the second approach, we also implant two trojans into one single model. Similar to the first approach, in order to trigger the backdoors, one needs to present different triggers to the infected model. However, different from the first approach, both backdoors are used to misclassify a trigger-inserted image into a single target class. In Figure~\ref{fig:2_demo}, we show a simple example, demonstrating the effect of the backdoors upon classification results.


To answer the question (4) and (7), in addition to training the same neural networks on the aforementioned two datasets, we trained learning models with more complicated neural architectures. To be specific, we trained 10 Conv + 5 MaxPooling on GTSRB and ResNet~\cite{he2016deep} on ImageNet (see Appendix for more details as well as other hyperparameters). Similar to the setup above, we also inserted trojan backdoors (\ie square trigger located at the bottom right with size $8\times8$ for GTSRB and firefox trigger located at top right with size $40\times40$ for ImageNet) to these trained models using BadNet and generate victim models. We again applied \sys and {\tt Neural Cleanse} to these victim learning models. In this way, we can compare their performance accordingly.

To answer the question (5) and (7), we first trained one neural network with the architectures of 6 Conv + 2 MaxPooling CNN on GTSRB. Then, we inserted a trojan backdoor (\ie square trigger located at the bottom right with size $8\times8$) into the model using BadNet. Again, we used both \sys and {\tt Neural Cleanse} to identify the existence of a trojan backdoor. But, instead of using the optimal hyperparameters manually identified (see details in Appendix), we subtly varied the corresponding hyperparameters for both techniques and observed the changes in their detection results. To be specific, for \sys, we selected two values for each of the hyperparameters $\lambda_1 \sim \lambda_6$ and then test all of their combinations. For {\tt Neural Cleanse}, we selected four values for its only tunable hyperparameter (\ie the weight of the penalty term that controls the size of the trigger). In Appendix, we list all the hyperparameters selected.

As is mentioned in Section~\S\ref{sec:background}, an alternative approach to implanting a trojan backdoor is Trojan Attack proposed in~\cite{Trojannn}. Therefore, we answer the question (6) and (7) as follows. First, we trained a neural network model using the dataset LFW because LFW is the dataset selected by {\tt Neural Cleanse} to demonstrate its defense against the Trojan Attack. With this clean model in hand, we then implanted a trojan backdoor to the model and thus generated the victim model using Trojan Attack~\cite{Trojannn} under its default setting~\cite{wang-2019-ieeesp}. Similar to other experiment designs above, in order to respond to question (7), we applied both \sys and {\tt Neural Cleanse} to examine the existence of trojan backdoors against the victim model and compared their performance in trojan detection. 

\subsection{Datasets and Evaluation Metrics}
The aforementioned experiments involve three datasets and we trained different neural architectures on these datasets. As a result, we first introduce these datasets. Since our designed experiments also involve the comparison between detection techniques, we further introduce the evaluation metrics for performance comparison. 

\vspace{-10pt}
\subsubsection{Datasets and Neural Architectures}
\para{LFW. }
Labeled Faces in the Wild~\cite{learned2016labeled} (LFW) is a database of face photographs with the dimensionality of $224\times224\times3$ designed for studying the problem of unconstrained face recognition. In this work, we follow the default setting of~\cite{Trojannn} and sample a subset of 2,622 samples, where each sample belongs to a distinct class. 

\para{GTSRB. }
German Traffic Sign Recognition Dataset~\cite{stallkamp2012man} (GTSRB) contains photos of traffic signs, which can be partitioned into 39,209 training and 12,630 testing samples. In this dataset, the photos of traffic signs have the dimensionality of $32 \times 32 \times 3$, and they can be categorized into 43 distinct classes. 

\para{ImageNet. }
ImageNet~\cite{deng2009imagenet} is one of the most widely used datasets for object classification and recognition. The whole dataset contains more than 14 million images across 20,000+ categories, in which each image has the dimensionality of $224 \times 224 \times 3$.  In this work, we use a subset of $50$ class, where each class has $400$ samples.

\vspace{-10pt}
\subsubsection{Evaluation Metrics}
\label{subsec:metric}

We introduce 2 sets of metrics, which measure the fidelity of the restored trigger as well as the correctness of backdoor detection. Here, we provide their definition below.

\para{Fidelity measure. }
Recall that a trojan detection approach needs to restore and visualize a trojan backdoor so that a security analyst could manually examine the reported trojan backdoor and then take further actions (\eg patching victim models). As a result, we define three fidelity measures -- precision, recall and $F_{1}$ score -- to quantify the quality of the restored backdoor. 

Given a trojan backdoor restored (\ie resorted trigger pertaining to the backdoor), the three measures are defined as follows. 
\begin{equation}
\begin{aligned}
    \text{precision} &= \frac{\|\bM \odot \bM _{t}\|_{1}} {\|\bM \|_{1}}, \ \text{recall} = \frac{\|\bM  \odot \bM _{t}\|_{1}}{\|\bM_{t}\|_{1}} \, , \\
    F_{1} &= 2 \cdot \frac{\text{precision} \cdot \text{recall}}{\text{precision} + \text{recall}} \, .
\label{eq:measure}
\end{aligned}
\end{equation}
Here, $\bM$ and $\bM _t$ represent the mask of the trigger restored and that of the ground-truth trigger, respectively. $\|\cdot\|_{1}$ denotes the $L_1$ norm. As we can see from the definition , the precision measure specifies the percentage of the restored trigger area truly overlapping with the ground-truth trigger area, whereas the recall measure describes the percentage of the ground-truth area correctly restored by a detection approach. $F_{1}$ score is the harmonic mean of precision and recall, and thus describes the overall quality of a restored trojan backdoor. It is not difficult to note that the higher these measures are, the better fidelity a restored trojan backdoor exhibits. 

\noindent{\bf Correctness measure. } 
This measure indicates the correctness of a trojan detection approach. In this work, we use 4 different symbols to represent its values -- \CIRCLE~(success detection), \RIGHTcircle~(success detection with errors), $\circleddash$~(incorrect trojan detection) and $\ocircle$~(failure detection). 

Given a learning model, \CIRCLE~indicates two situations. First, it means a detection approach could successfully pinpoint the trojan backdoors intentionally implanted in an infected model, and does not mistakenly report the existence of additional trojan backdoors capable of misclassifying a trigger-inserted image into an uninfected class. Second, for a clean learning model without a trojan backdoor implanted, it indicates a detection approach correctly asserts the nonexistence of backdoors. 

Similar to \CIRCLE, \RIGHTcircle~also represents the successful identification of a trojan backdoor intentionally inserted. Differently, it, however, indicates a situation where a detection approach also mistakenly pinpoints the existence of additional backdoors. For example, a trojan backdoor is inserted into a victim learning model, which misclassifies a trigger-inserted image into target class A. Using a trojan detection approach, we can successfully determine the existence of a trojan backdoor misleading prediction results to the target class A. But, along with this correct identification, it also falsely reports an additional trojan which misclassifies a trigger-inserted image into an uninfected class B. 

With respect to $\circleddash$, it means that, given a learning model with a backdoor inserted, a detection approach reports the existence of a backdoor. However, different from the situations above, it fails to tie the detected trojan backdoor to the correct infected class. Again, take the aforementioned case for example. Using a trojan detection approach with the correctness of $\circleddash$, we deem a learning model contains a backdoor or, in other words, track down an incorrect trojan backdoor, not the one intentionally inserted. Unfortunately, we, however, cannot correctly identify the trojan backdoor pertaining to the target class A. Regarding $\ocircle$, it simply indicates a detection approach (1) fails to deem a victim learning model contains a trojan backdoor intentionally inserted or (2) mistakenly reports the existence of backdoor when the model is actually clean, carrying no trojan backdoors.


As is discussed in Section~\S~\ref{sec:background}, given a learning model, a model user needs a technical approach to \ding{182} correctly assert the (non-)existence of a manufactured trojan backdoor and \ding{183} -- if a trojan exists -- restore the corresponding trigger in a high-fidelity fashion. With a technique like this in hand, she could manually examine a learning model and take actions accordingly. As a result, a trojan detection approach is more favorable if its detection correctness is marked as \CIRCLE~or \RIGHTcircle. On the contrary, a detection approach is less useful if its correctness is marked as $\circleddash$ or $\ocircle$. This is simply because $\ocircle$ implies the complete failure of a detection approach whereas $\circleddash$ indicates a false detection which might mislead a security analyst into taking wrong actions. 

\subsection{Experiment Results}

\newcommand{\tabincell}[2]{\begin{tabular}{@{}#1@{}}#2\end{tabular}}

\begin{table*}[hbtp]
\begin{subtable}[t]{0.9\textwidth}
\small
\centering
\begin{tabular}{c|c|c|c|c|c|c|c|c|c|c|c}
\Xhline{1.05pt}
& \multirow{3}{*}{\tabincell{c}{\\ Shape}}    & \multirow{3}{*}{\tabincell{c}{\\ Position}}                                               & \multirow{3}{*}{\tabincell{c}{\\ Size}} & \multicolumn{6}{c|}{Fidelity Measure}                                                   & \multicolumn{2}{c}{Correctness Measure}            \\ \cline{5-12} 
&                          &                                                                         &                       & \multicolumn{2}{c|}{Precision} & \multicolumn{2}{c|}{Recall} & \multicolumn{2}{c|}{F1} & {\tt NCleanse} & \multirow{2}{*}{\tabincell{c}{\\ \sys}}  \\ \cline{5-10}
                        &  &                                                                         &                       & {\tt NCleanse}         & \sys          & {\tt NCleanse}        & \sys         & {\tt NCleanse}     & \sys       &                          &                       \\ \Xhline{1.05pt}
Clean  & N/A & N/A & N/A & N/A & N/A & N/A & N/A & N/A & N/A & $\ocircle$ & \CIRCLE \\ \Xhline{1.05pt}
\multirow{20}{*}{\begin{tabular}[c]{@{}c@{}}Victim\\ model\end{tabular}} 
& \multirow{10}{*}{\begin{tabular}[c]{@{}c@{}}White\\ Square\end{tabular}}  & \multirow{5}{*}{\begin{tabular}[c]{@{}c@{}}Top\\ Left\end{tabular}} & $6\times6$  & 0.449 & \textbf{1.000} & \textbf{0.611} & 0.426 & 0.518 & \textbf{0.597} &\CIRCLE  & \CIRCLE \\ \cline{4-12} 
                        &  &                              &$8\times8$  & 0.544 & \textbf{0.964} & 0.672 & \textbf{0.703} & 0.601 & \textbf{0.813}    & $\ocircle$ & \CIRCLE  \\ \cline{4-12} 
                        &  &                              & $10\times10$  & 0.718 & \textbf{1.000} & 0.518 & \textbf{0.580} & 0.596 & \textbf{0.734} & $\ocircle$ & \CIRCLE \\ \cline{4-12} 
                        &  &                              & $12\times12$  & 0.540 & \textbf{1.000} & 0.375 & \textbf{0.431} & 0.443 & \textbf{0.602} & $\ocircle$ & \CIRCLE   \\ \cline{4-12} 
                        &  &                              & $14\times14$  & 0.561 & \textbf{0.816} & 0.281 & \textbf{0.980} & 0.374 & \textbf{0.890} & $\ocircle$ & \CIRCLE \\ \Xcline{3-12}{1.05pt} 
                        &  & \multirow{5}{*}{\begin{tabular}[c]{@{}c@{}}Bottom\\ Right\end{tabular}} & $6\times6$  & 0.186 & \textbf{0.870} & 0.750 & \textbf{0.972} & 0.298 & \textbf{0.918} & $\ocircle$ & \CIRCLE \\ \cline{4-12} 
                        &  &                              & $8\times8$  & 0.578 & \textbf{0.948} & 0.578 & \textbf{0.984} & 0.578 & \textbf{0.966} & $\ocircle$ & \RIGHTcircle \\ \cline{4-12} 
                        &  &                              & $10\times10$ & 0.671 & \textbf{0.966} & 0.530 & \textbf{0.770} & 0.592 & \textbf{0.857} & $\ocircle$ & \RIGHTcircle \\ \cline{4-12} 
                        &  &                              & $12\times12$ & 0.643 & \textbf{0.983} & 0.500 & \textbf{0.500} & 0.563 & \textbf{0.663} & $\ocircle$ & \CIRCLE\\ \cline{4-12} 
                        &  &                              & $14\times14$ & 0.812 & \textbf{0.980} & \textbf{0.571} & 0.541 & 0.671 & \textbf{0.697} & $\ocircle$ & \RIGHTcircle \\ \Xcline{2-12}{1.05pt}
&\multirow{10}{*}{\begin{tabular}[c]{@{}c@{}}Firefox\\ Logo\end{tabular}} & \multirow{5}{*}{\begin{tabular}[c]{@{}c@{}}Top\\ Right\end{tabular}}    & $6\times6$  & 0.458 & \textbf{1.000} & 0.177 & \textbf{0.452} & 0.256 & \textbf{0.622} & $\ocircle$ & \CIRCLE\\ \cline{4-12} 
                        &  &                              & $8\times8$  & 0.448 & \textbf{1.000} & \textbf{0.732} & 0.285 & \textbf{0.556} & 0.443  & $\circleddash$  & \RIGHTcircle\\ \cline{4-12} 
                        &  &                              & $10\times10$ & 0.882 & \textbf{0.971} & 0.233 & \textbf{0.337} & 0.369 & \textbf{0.500} & \RIGHTcircle  & \RIGHTcircle \\ \cline{4-12} 
                        &  &                              & $12\times12$ & \textbf{0.963} & 0.667 & 0.193 & \textbf{0.274} & 0.321 & \textbf{0.388} & \RIGHTcircle  & \RIGHTcircle  \\ \cline{4-12} 
                        &  &                              & $14\times14$ & 0.477 & \textbf{1.000} & \textbf{0.425} & 0.275 & \textbf{0.450} & 0.432 & $\ocircle$    & \RIGHTcircle  \\ \Xcline{3-12}{1.05pt} 
                        &  & \multirow{5}{*}{\begin{tabular}[c]{@{}c@{}}Bottom\\ Left\end{tabular}}  & $6\times6$     & 0.052 & \textbf{0.952} & 0.210 & \textbf{0.355} & 0.083 & \textbf{0.517} & $\ocircle$ &\RIGHTcircle   \\ \cline{4-12} 
                        &  &                             & $8\times8$     & \textbf{0.974} & 0.958 & 0.309 & \textbf{0.398} & 0.461 & \textbf{0.563}  & \RIGHTcircle & \RIGHTcircle \\ \cline{4-12} 
                        &  &                            & $10\times10$    & 0.493 & \textbf{0.873} & \textbf{0.575} & 0.518 & 0.531 & \textbf{0.650} & $\ocircle$     & \RIGHTcircle \\ \cline{4-12} 
                        &  &                            & $12\times12$    & 0.871 & \textbf{0.963} & 0.300 & \textbf{0.393} & 0.446 & \textbf{0.558} & $\ocircle$     & \CIRCLE         \\ \cline{4-12} 
                        &  &                         & $14\times14$      & \textbf{0.600} & 0.554 & 0.144 & \textbf{0.262} & 0.232 & \textbf{0.356}   & \RIGHTcircle    & \RIGHTcircle    \\ \Xhline{1.05pt}
\end{tabular}
\caption{Trojan Fidelity and End-to-end Detection Results on GTSRB.}
\label{tab:pri_gtsrb}
\end{subtable}
\\
\begin{subtable}[t]{0.9\textwidth}
\small
\centering
\begin{tabular}{c|c|c|c|c|c|c|c|c|c|c|c}
\Xhline{1.05pt}
 &\multirow{3}{*}{\tabincell{c}{\\ Shape}}    & \multirow{3}{*}{\tabincell{c}{\\ Position}}  & \multirow{3}{*}{\tabincell{c}{\\ Size}} & \multicolumn{6}{c|}{Fidelity Measure}  &\multicolumn{2}{c}{Correctness Measure}            \\ \cline{5-12} 
 &   &                                                   &                       & \multicolumn{2}{c|}{Precision} & \multicolumn{2}{c|}{Recall} & \multicolumn{2}{c|}{F1} &{\tt NCleanse} & \multirow{2}{*}{\tabincell{c}{\\ \sys}}  \\ \cline{5-10}
                          &    &                                                                     &                       & {\tt NCleanse}          & \sys          & {\tt NCleanse}        & \sys         & {\tt NCleanse}     & \sys       &                          &                       \\ \Xhline{1.05pt}
Clean &N/A &N/A &N/A & N/A & N/A & N/A & N/A & N/A & N/A & \CIRCLE & \CIRCLE \\ \Xhline{1.05pt}
\multirow{20}{*}{\begin{tabular}[c]{@{}c@{}}Victim\\ model\end{tabular}} & \multirow{10}{*}{\begin{tabular}[c]{@{}c@{}}Black\\ Square\end{tabular}}  & \multirow{5}{*}{\begin{tabular}[c]{@{}c@{}}Top\\ Left\end{tabular}} & $20\times20$  & \textbf{1.000} & 0.852 & 0.068 & \textbf{0.448} & 0.126 & \textbf{0.587} &$\ocircle$  & \CIRCLE \\ \cline{4-12}
                        &  &                              &$40\times40$ & 0.598 & \textbf{0.889} & 0.033 & \textbf{0.309} & \textbf{0.616} & 0.459   & $\ocircle$ & \RIGHTcircle \\ \cline{4-12} 
                        &  &                              & $60\times60$  & 0.670 & \textbf{0.997} & 0.043 & \textbf{0.932} & 0.080 & \textbf{0.963}  & \CIRCLE & \RIGHTcircle \\ \cline{4-12} 
                        &  &                              & $80\times80$  & 0.811 & \textbf{1.000} & 0.047 & \textbf{0.663} & 0.089 & \textbf{0.797} & $\ocircle$ & \RIGHTcircle   \\ \cline{4-12} 
                        &  &                              & $100\times100$  & \textbf{1.000} & 0.648 & 0.038 & \textbf{0.411} & 0.073 & \textbf{0.503}  & $\ocircle$ & \RIGHTcircle \\ \Xcline{3-12}{1.05pt} 
                        &  & \multirow{5}{*}{\begin{tabular}[c]{@{}c@{}}Bottom\\ Right\end{tabular}} & $20\times20$ & 0.394 & \textbf{1.000} & 0.033 & \textbf{0.285} & 0.060 & \textbf{0.444}  & $\ocircle$ & \RIGHTcircle \\ \cline{4-12} 
                        &  &                              & $40\times40$ & 0.731 & \textbf{0.895} & 0.066 & \textbf{0.323} & 0.121 & \textbf{0.475} & $\ocircle$ & \RIGHTcircle \\ \cline{4-12} 
                        &  &                              & $60\times60$ & 0.228 & \textbf{0.766} & 0.026 & \textbf{0.931} & 0.047 & \textbf{0.841}  & $\ocircle$ & \RIGHTcircle \\ \cline{4-12} 
                        &  &                              & $80\times80$ & 0.458 & \textbf{1.000} & 0.023 & \textbf{0.552} & 0.044 & \textbf{0.711} & $\ocircle$ & \RIGHTcircle\\ \cline{4-12} 
                        &  &                              & $100\times100$  & 0.000 & \textbf{0.913} & 0.000 & \textbf{0.438} & NaN & \textbf{0.592}  & $\ocircle$ & \RIGHTcircle \\ \Xcline{2-12}{1.05pt}
& \multirow{10}{*}{\begin{tabular}[c]{@{}c@{}}Firefox\\ Logo\end{tabular}} & \multirow{5}{*}{\begin{tabular}[c]{@{}c@{}}Top\\ Right\end{tabular}}    & $20\times20$ & 0.061 & \textbf{0.398} & 0.011 & \textbf{0.168} & 0.019 & \textbf{0.237} & $\ocircle$ & \RIGHTcircle\\ \cline{4-12} 
                        &  &                              & $40\times40$ & 0.664 & \textbf{0.752} & 0.072 & \textbf{0.197} & 0.129 & \textbf{0.313} & $\ocircle$  & \CIRCLE\\ \cline{4-12} 
                        &  &                              & $60\times60$ & \textbf{0.898} & 0.779 & 0.082 & \textbf{0.118} & 0.150 & \textbf{0.205} & \CIRCLE  & \RIGHTcircle \\ \cline{4-12} 
                        &  &                              & $80\times80$ & 0.312 & \textbf{0.774} & 0.037 & \textbf{0.141} & 0.066 & \textbf{0.238}  & $\ocircle$  & \RIGHTcircle \\ \cline{4-12} 
                        &  &                              & $100\times100$& 0.902 & \textbf{0.925} & 0.056 & \textbf{0.105} & 0.106 & \textbf{0.189} & $\ocircle$    & \RIGHTcircle \\ \Xcline{3-12}{1.05pt} 
                        &  & \multirow{5}{*}{\begin{tabular}[c]{@{}c@{}}Bottom\\ Left\end{tabular}}  & $20\times20$   & \textbf{1.000} & 0.048 & \textbf{0.135} & 0.011 & \textbf{0.239} & 0.018  & $\ocircle$ &\RIGHTcircle  \\ \cline{4-12} 
                        &  &                          & $40\times40$     & 0.600 & \textbf{0.917} & 0.046 & \textbf{0.271} & 0.086 & \textbf{0.418}   & $\ocircle$ & \CIRCLE \\ \cline{4-12} 
                        &  &                          & $60\times60$ & 0.803 & \textbf{0.929} & 0.081 & \textbf{0.147} & 0.147 & \textbf{0.254} & $\ocircle$     & \RIGHTcircle \\ \cline{4-12} 
                        &  &                          & $80\times80$ & 0.000 & \textbf{0.889} & 0.000 & \textbf{0.044} & NaN & \textbf{0.083} & $\ocircle$     & \RIGHTcircle  \\ \cline{4-12} 
                        &  &                          & $100\times100$      & 0.000 & \textbf{0.951} & 0.000 & \textbf{0.089} & NaN & \textbf{0.163}  & $\ocircle$    & \CIRCLE    \\ \Xhline{1.05pt}
\end{tabular}
\caption{Trojan Fidelity and End-to-end Detection Results on ImageNet.}
\label{tab:pri_imagenet}
\end{subtable}
\caption{Performance comparison between {\tt Neural Cleanse}, indicated by {\tt NCleanse} and \sys on Trojan Fidelity and End-to-end Detection. Note that N/A stands for ``Not- Available''. NaN represents ``Not-A-Number''.} 
\label{tab:pri}
\vspace{-20pt}
\end{table*}

In the following, we show and summarize the experiment results. In addition, we discuss and analyze the reasons for our findings.
\begin{figure}[htbp]
    \begin{subfigure}[b]{0.23\textwidth}
        \includegraphics[width=\textwidth]{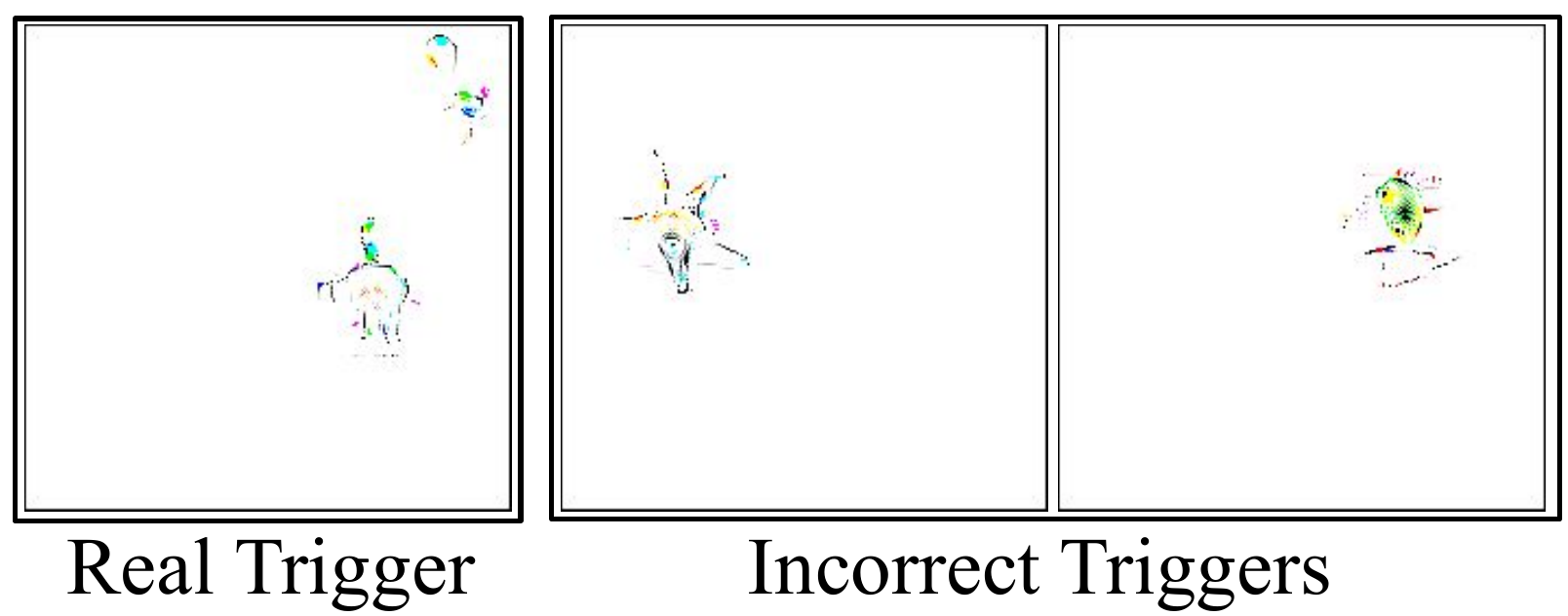}
        \caption{{\tt Neural Cleanse}.}
        \label{subfig:cleanse_fp}
    \end{subfigure}
    \hspace{3pt}
    \begin{subfigure}[b]{0.23\textwidth}
        \includegraphics[width=\textwidth]{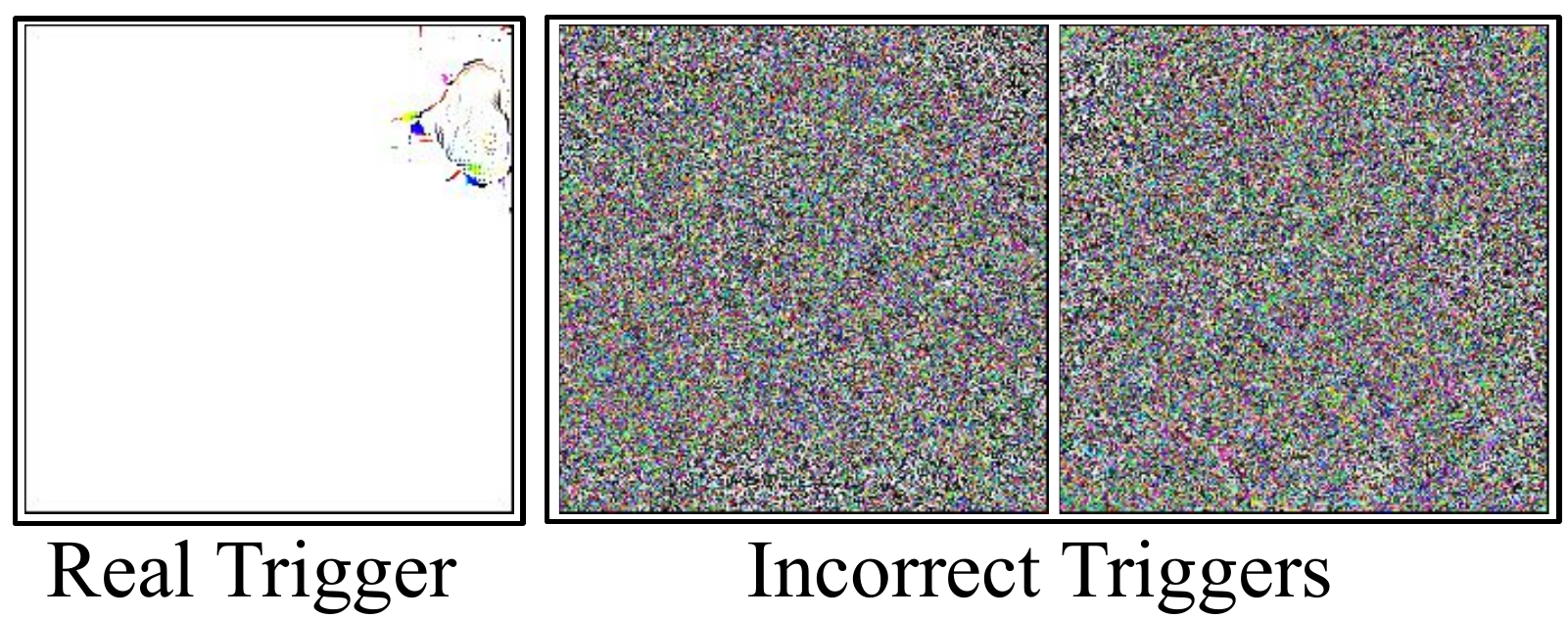}
        \caption{\sys.}
        \label{subfig:our_fp}
    \end{subfigure}
    \caption{Real triggers \& incorrect triggers detected by {\tt Neural Cleanse} and \sys in ImageNet. Note that for better visualization quality, we reverse the color of the restored trojans from ImageNet dataset.}
    \label{fig:false_positive}
    \vspace{-8pt}
\end{figure}
\subsubsection{Primary Experiment Results}

In Table~\ref{tab:pri}, we show some of the comparison results. They indicate the performance of our technique as well as that of {\tt Neural Cleanse} observed at the situations where we change the size, shape, and position of a trigger pertaining to a trojan backdoor. 

As we can observe from Table~\ref{tab:pri}, for learning models that enclose a trojan backdoor with a different size, in a different shape or at a different location, {\tt Neural Cleanse} oftentimes cannot point out the trojan existence (indicated by the symbol $\ocircle$). Even if it sometimes points out trojan existence, in one of the cases, it fails to pinpoint the trigger truly inserted (indicated by the symbol $\circleddash$). For a clean model carrying no trojan backdoor, we also discover {\tt Neural Cleanse} mistakenly reports the model is implanted with a trojan backdoor (represented by symbol $\ocircle$). There are two major reasons behind these results. 

First, as is discussed in Section~\S\ref{sec:tech}, a trigger pertaining to a trojan backdoor is a special adversarial sample. In~\cite{wang-2019-ieeesp}, Wang \etal designed {\tt Neural Cleanse} to search that trigger in a specific adversarial subspace under the guidance of an optimization objective. With the variation in trigger shapes, sizes and positions, the adversarial subspace got varied. In different adversarial subspaces, the total number of adversarial samples might vary, influencing the stability of {\tt Neural Cleanse} in trojan detection. Intuition suggests the more adversarial samples there are in an adversarial space, the less likely it will be for an optimization technique to pinpoint the target trigger.

Second, regardless of whether a backdoor exists in a learning model, the optimization-based technique could always find a local optimal solution for a corresponding optimization problem and thus deem it as a reasonably good local optimum (\ie a trojan backdoor). In~\cite{wang-2019-ieeesp}, Wang \etal proposes a technique to distinguish local optima (incorrect triggers) from reasonably good local optima (implanted trojan backdoors). Technically speaking, they design a distinguishing metric and utilize a anomaly detection technique to determine if a local optimum represents the trojan backdoor intentionally inserted. However, as we depict in Figure~\ref{fig:false_positive}, the incorrect triggers, and target backdoors have less distinguishability between each other (\ie both have relatively dense pixels grouping together).
\begin{figure}[htbp]
    \centering
    \begin{subfigure}[b]{0.45\textwidth}
        \includegraphics[width=\textwidth]{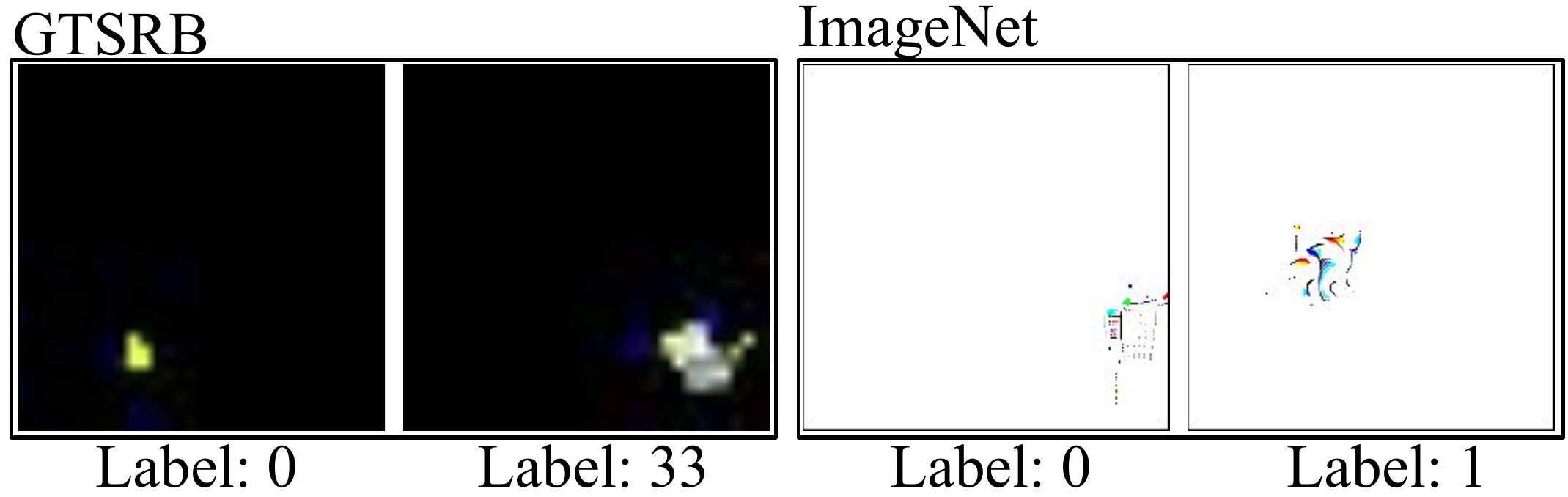}
        \caption{{\tt Neural Cleanse}.}
        \label{subfig:sp_2_2}
    \end{subfigure}\\
    \begin{subfigure}[b]{0.45\textwidth}
        \includegraphics[width=\textwidth]{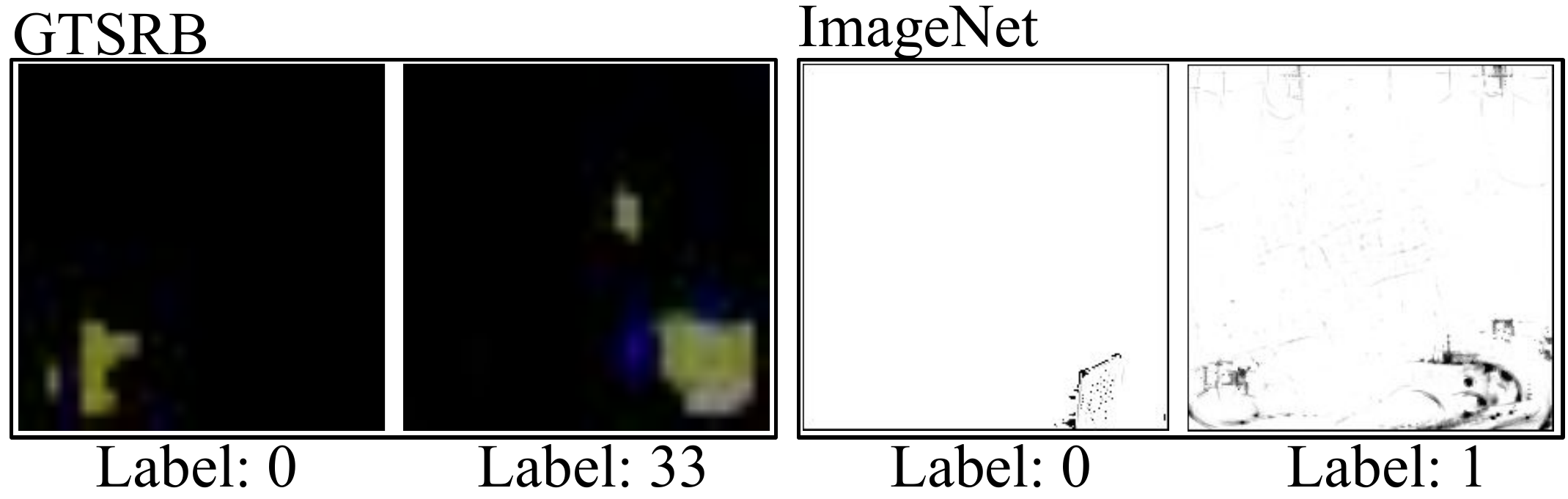}
        \caption{\sys.}
        \label{subfig:our_2_2}
    \end{subfigure}
    \vspace{-5pt}
    \caption{Triggers detected under the setting where two backdoors are tied to two different classes in one victim model. In end-to-end detection, {\tt Neural Cleanse} detects the trojan with some false positives on GTSRB, and reports three wrongly-detected trojans on ImageNet. \sys correctly reports the trojaned label without any false positives on both datasets.}
    \label{fig:2_trojan_2_label}
    \vspace{-5pt}
\end{figure}

In comparison with the extremely poor performance {\tt Neural Cleanse} exhibits in the setting of trigger shape/size/location variations, our technique demonstrates a significant improvement in trojan backdoor detection\footnote{Table~\ref{tab:pri} demonstrates that both approaches are relatively robust to trojan colors.}. As is shown in Table~\ref{tab:pri} and Figure~\ref{fig:false_positive} (more examples of the restored triggers are shown in the Appendix), for all settings, our technique could accurately point out the existence or non-existence of a trojan backdoor, and the incorrect triggers and target backdoors have more obvious distinguishability. On the one hand, this is because our design reduces the adversarial subspace which eases the search of triggers intentionally inserted. On the other hand, this is due to the fact that our technique integrates a more effective anomaly detection mechanism (\ie the new metric designed in~\S\ref{sec:tech}) which could distinguish triggers from incorrect triggers. From Table~\ref{tab:pri}, we also note that, given a learning model with one implanted backdoor, our technique typically identifies additional trojan backdoors pertaining to non-target classes. However, as is discussed in Section~\S\ref{subsec:metric}, these false identifications do not jeopardize the effectiveness of our proposed technique because (1) the most important goal of this work is to determine the existence of the trojan backdoor (defined in Section~\S\ref{sec:background}), and our approach can always pinpoint the trojan intentionally inserted and does not falsely report the trojan existence for clean learning models. (2) these false identifications do not influence further patching actions in that an infected model will be patched using the correct trigger, and patching additional non-intentionally inserted trojans does not affect the accuracy of the patched model.

Going beyond the performance comparison from the perspective of detection correctness, we also show the performance of our approach from the fidelity aspect. As is illustrated in Table~\ref{tab:pri}, overall, our approach demonstrates higher measures in precision, recall, and F-score than {\tt Neural Cleanse}. This indicates our approach could restore a trojan backdoor (\ie the trigger pertaining to the backdoor) with higher fidelity. As is discussed in Section~\S\ref{sec:background}, this capability is critical because it could better facilitate a model user to examine trojan existence and thus take actions.
\begin{figure}[t]
    \centering
    \begin{subfigure}[b]{0.18\textwidth}
        \includegraphics[width=\textwidth]{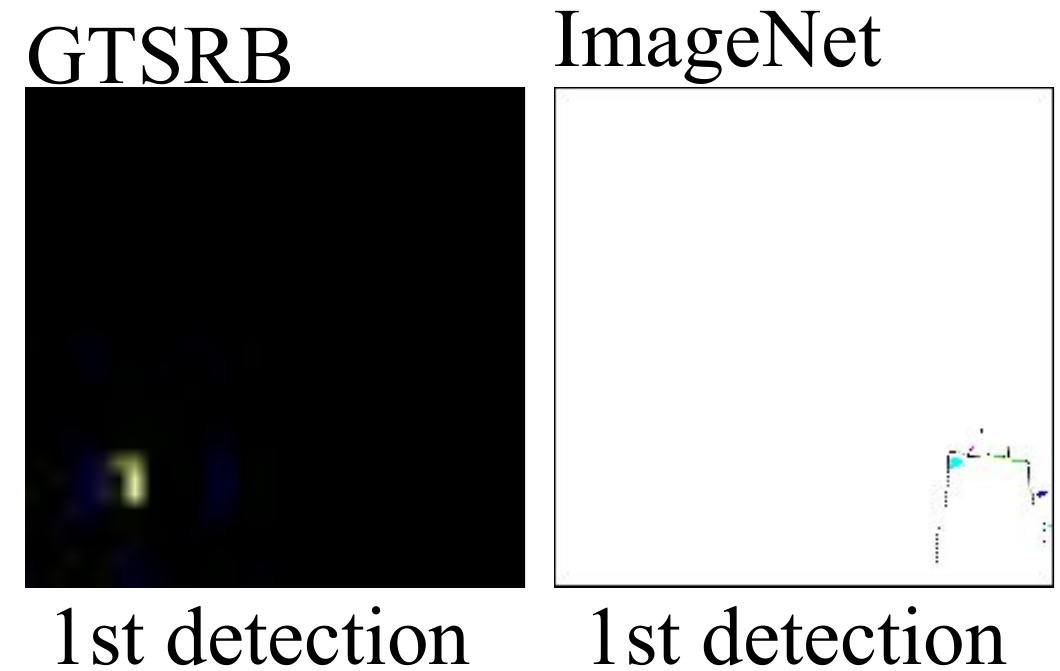}
        \caption{{\tt Neural Cleanse}.}
        \label{subfig:2_1_sp}
    \end{subfigure}
    \begin{subfigure}[b]{0.27\textwidth}
        \includegraphics[width=\textwidth]{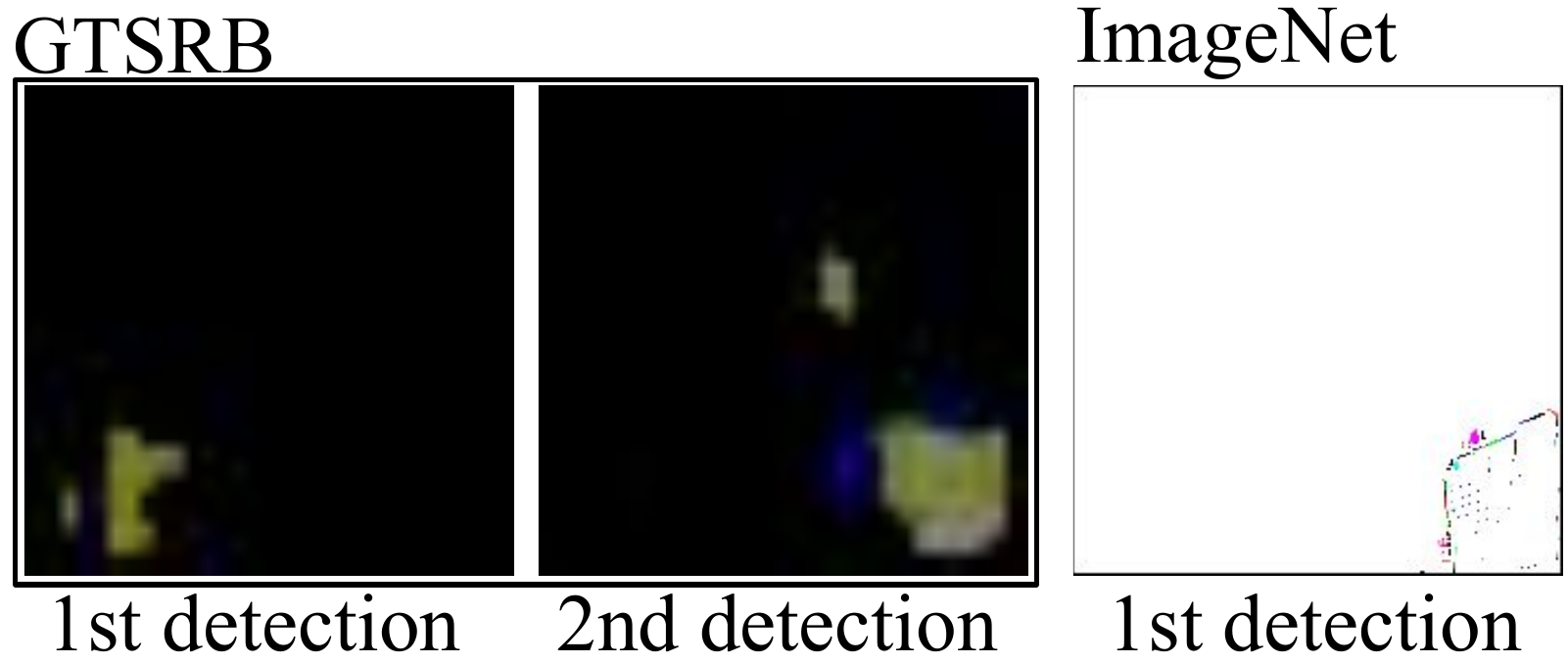}
        \caption{\sys.}
        \label{subfig:2_1_our}
    \end{subfigure}
    \vspace{-10pt}
    \caption{Triggers detected under the setting where two backdoors are tied to one class in one victim model.}
    \label{fig:2_trojan_1_label}
    \vspace{-10pt}
\end{figure}

Last, as is shown in Table~\ref{tab:pri}, while we perform trojan detection on learning models taking input in different dimensionalities (ImageNet with $224 \times 224$ and GTSRB with $32 \times 32$), for both {\tt Neural Cleanse} and our approach, the difference in input dimensionality has less impact upon their backdoor detection. This indicates that we can expect consistent detection performance if we utilize these approaches to detect trojan backdoors against models trained on other data sets with different input dimensionalities.

\begin{figure*}[htbp]
    \centering
    \begin{subfigure}[b]{0.16\textwidth}
        \includegraphics[width=\textwidth]{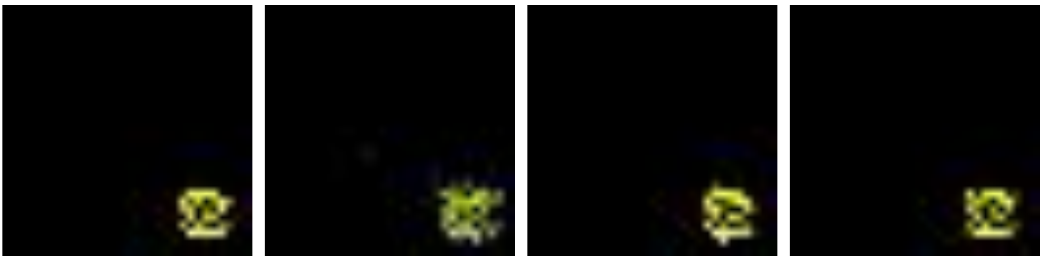}
        \caption{{\tt Neural Cleanse}.}
        \label{subfig:hyper_sp}
    \end{subfigure}\hspace{6pt}
    \begin{subfigure}[b]{0.64\textwidth}
        \includegraphics[width=\textwidth]{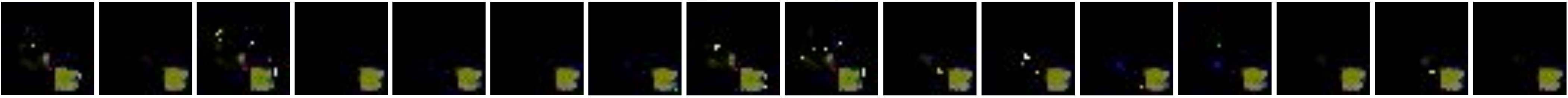}
        \caption{\sys.}
        \label{subfig:hyper_our}
    \end{subfigure}
    \vspace{-10pt}
    \caption{Triggers detected under the same setting using different hyperparameters.}
    \label{fig:hyper}
    \vspace{-5pt}
\end{figure*}

\vspace{-5pt}
\subsubsection{Other Experiment Results} 

As is mentioned above, we also design an experiment inserting two backdoors into one single victim learning model and tying them to different target classes. In Figure~\ref{fig:2_trojan_2_label}, we show the experiment results observed under this setting. As we can observe from Figure~\ref{subfig:our_2_2}, our technique could not only successfully point out the existence of both trojan backdoors but accurately tie them to the correct target labels as well. In contrast, {\tt Neural Cleanse} either fails to point out the existence of trojan backdoor (for GTSRB) or only deems incorrect trojans as true backdoors (for ImageNet). We believe this observation results from the fact that our technique not only shrinks the adversarial subspace to better resolve the true injected trojans but also integrates a better anomaly detection metric than {\tt Neural Cleanse} to distinguish falsely identified backdoors (\ie incorrect triggers) from the true backdoors. 

\begin{figure}[htbp]
    \centering
    \begin{subfigure}[b]{0.45\textwidth}
        \includegraphics[width=\textwidth]{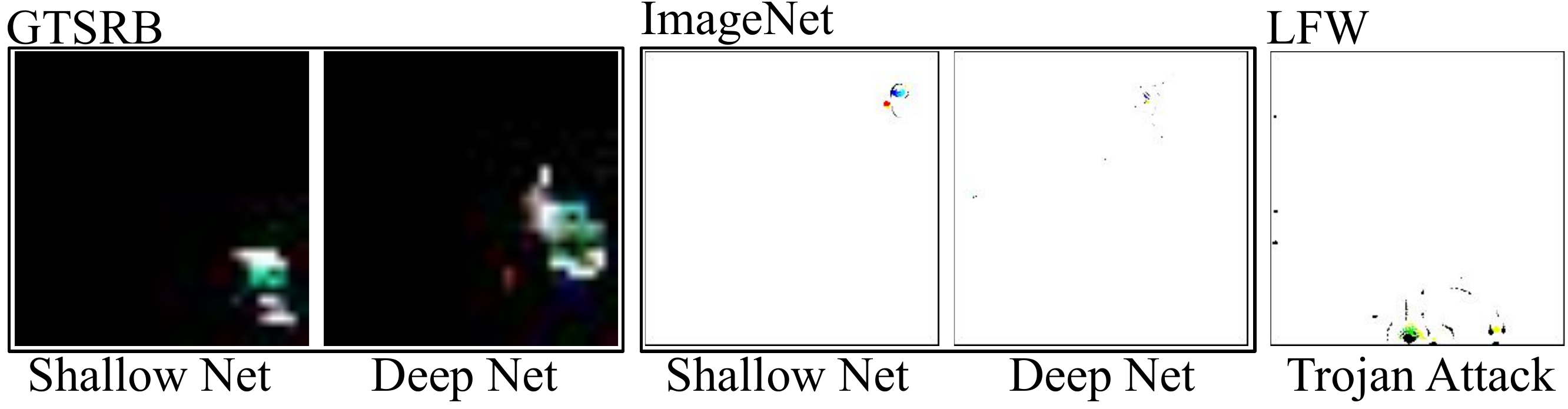}
        \caption{{\tt Neural Cleanse}.}
        \label{subfig:comp_sp}
    \end{subfigure}\\
    \begin{subfigure}[b]{0.45\textwidth}
        \includegraphics[width=\textwidth]{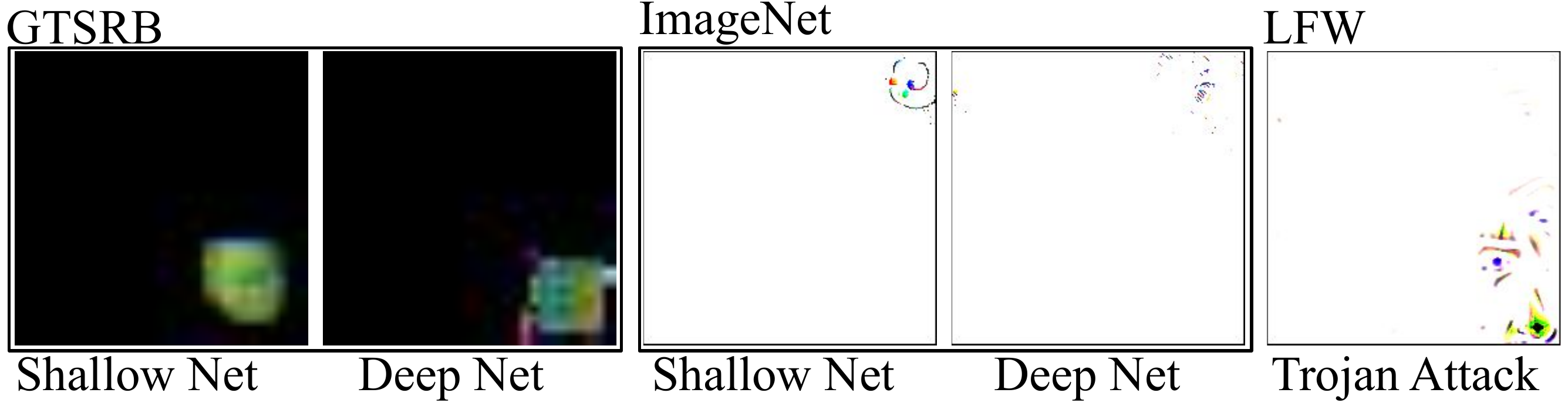}
        \caption{\sys.}
        \label{subfig:comp_our}
    \end{subfigure}
    \vspace{-5pt}
    \caption{Triggers detected under the alternative settings. The correctness measure results of the deep nets are consistent with those of the shallow networks shown in Table~\ref{tab:pri}. In end-to-end detection, both {\tt Neural Cleanse} and \sys manage to detect the correct trigger without any false positives. Note that we also reverse the color of the restored triggers from LFW dataset.}
    \label{fig:complexity}
    \vspace{-5pt}
\end{figure}

In addition to showing the performance under the situation where two backdoors are inserted and tied to two different classes, Figure~\ref{fig:2_trojan_1_label} illustrates the experimental results obtained from the alternative setting where two backdoors are inserted and attached to the same class. As we can observe from Figure~\ref{subfig:2_1_sp} and \ref{subfig:2_1_our}, both {\tt Neural Cleanse} and our approach point out the existence of trojan backdoors. However, as is depicted in Figure~\ref{fig:2_trojan_1_label}, both of these approaches can only restore the trigger tied to one trojan backdoor but fail to recover the trigger tied to the other. We believe this is presumably because in the adversarial subspace one backdoor is easier to be identified than the other when we perform a search under the guidance of an objective function. While this might influence an analyst to examine the backdoor and maybe patch both triggers, it does not harm the effectiveness of both detection approaches in pointing out the existence of trojan backdoors.

In this work, we have utilized the \emph{unlearning} method introduced in~\cite{wang-2019-ieeesp} to patch the victim model with the expectation of seeing the successful restoration of the other trigger. In Figure~\ref{fig:2_trojan_1_label}, we show this result. As we can observe, after patching the backdoor identified by both approaches and then applying both approaches to the patched models, {\tt Neural Cleanse} still fails to find the missing triggers. However, we surprisingly find our approach could accurately pinpoint the missing trigger for the model originally trained on GTSRB but fails to identify the missing trigger pertaining to the model originally trained on ImageNet. This indicates that for low dimensional inputs, restoring a high fidelity trigger is enough for patching the infected model even simply applying \emph{unlearning}. However, for high dimensional inputs, we may need both a high fidelity restored trigger and an advanced patching method to patch an infected model due to the extremely large adversarial subspace of high dimensional inputs.


In addition to the aforementioned observations, we discover that, in both of the settings above, our technique always restores trojan backdoors with higher fidelity than {\tt Neural Cleanse} (see Figure~\ref{fig:2_trojan_1_label}). This observation aligns with our experiment results observed in the experiment above and implies that our approach could potentially provide an analyst with better capability in manually examining trojan backdoors and thus better patching a victim learning model.

As is described in Section~\S\ref{sec:tech}, both our technique and {\tt Neural Cleanse} formalize trojan detection as an optimization problem. In order to solve the optimization problem, objective functions are defined. In these objective functions, both our technique and {\tt Neural Cleanse} introduce multiple hyperparameters -- 6 hyperparameters for our approach and 1 hyperparameter for {\tt Neural Cleanse}. In Figure~\ref{subfig:hyper_sp} and \ref{subfig:hyper_our}, we illustrate some triggers restored under different hyperparameters. As we can observe, for our proposed technique and {\tt Neural Cleanse}, the subtle variation in hyperparameters has nearly no influence upon the trojan detection. This is because both approaches adopt the hyperparameter augmentation introduced in Section~\S\ref{sec:tech}. The results indicate we can expect the stable detection accuracy and restoration fidelity even if the hyperparameters are subtly varied. This is a critical characteristic because model users do not need to overly worry to set very precise hyperparameters in order to obtain the optimal detection performance. 

The experiment results discussed above are all collected from the settings where the learning models are trained with relatively shallow neural architectures and the trojan backdoors are inserted through a relatively simple method (\ie BadNet). In Figure~\ref{fig:complexity}, we show experimental results gathered from alternative settings. 

First, we can observe, when a victim model encloses a trojan backdoor inserted by a more advanced attack technique, both our approach and {\tt Neural Cleanse} could correctly identify the existence of a trojan backdoor and tie it to the right target class without mistakenly reporting additional backdoors. This indicates both approaches are robust against the way to implant trojan backdoors. However, we also note the trigger restored by both approaches have a relatively low fidelity compared to the trigger restored from the BadNet attack. We believe this is because the trigger inserted by Trojan Attack~\cite{Trojannn} has a higher complexity and, through the optimization objective designed by both approaches, it is more difficult to completely recover the trigger intentionally inserted.

Second, we can observe, when victim models are trained with a more complicated architecture, both our approach and {\tt Neural Cleanse} could correctly pinpoint the existence of trojan backdoors. This indicates that the complexity of target models does not influence the detection correctness, although more complicated learning models imply a more complicated search space, making trigger search more difficult. With this observation, we can conclude the performance variation of both approaches is not strongly tied to the change of model complexity. 


\section{Conclusion}
\label{sec:conclusion}

Given a target learning model, this work shows that, even for the state-of-the-art trojan backdoor detection technique, it is still difficult to accurately point out the existence of a trojan backdoor without false alarms or failure identification, particularly when the trigger size, shape, and its presentation location vary. Inspired by this, we propose a new technical approach. Technically, it first formalizes the detection of trojan backdoors as an optimization problem and identifies a set of candidate triggers by resolving this optimization problem. Then, it defines a new measure to quantify these candidate triggers and uses an anomaly detection approach to distinguish real triggers from incorrect triggers in an infected model and eliminate false alarms in an clean model. Following this design, we implement \sys and show that our technical approach can not only accurately point out the existence of trojan backdoors but more importantly restore the presentation of trojan backdoors. Thus, we conclude that an optimization-based technical approach can significantly escalate the accuracy of trojan backdoor detection and improve the fidelity of the resorted backdoors.

\bibliographystyle{ACM-Reference-Format}
\bibliography{ref}

\newpage
\appendix
\section{Complete Optimization Function}
Putting Equation~\eqref{eq:1} and four regularization terms introduced in Section~\S\ref{sec:tech} together, the proposed optimization objective function is as following:
\begin{equation}
\label{eq:8}
\begin{aligned}
    \text{argmin} \sum_{i = 1}^{n} L(\bx_i,\bM, \bd) +  R_1 +  R_2 + R_3 + R_4, \\
\end{aligned}
\end{equation}
where
\begin{equation}
\label{eq:9}
\begin{aligned}
    L(\bx_i,\bM, \bd) &= L(f(\bx_i \odot (\mathbf{1} - \bM ) + \bM \odot \bd), y_t) \, , \\
    R_1(\bM, \bd) &= \lambda_1\cdot R_{\text{elastic}}(\bM) + \lambda_2 \cdot R_{\text{elastic}}(\bd') \, , \\
    R_2(\bM, \bd) &= \lambda_3 \cdot R_{\text{smooth}}(\bM) + \lambda_4 \cdot R_{\text{smooth}}(\bd') \, , \\
    R_3(\bM, \bd, \bx_i) &= \lambda_5 \cdot L(f(\bx_i \odot (\mathbf{1} - \bM )), y_{t'_i}) \, , \\
    R_4(\bM, \bd) & = \lambda_6 \cdot L(f(\bM \odot \bd, y_t)  \, , \, \bd' = (\mathbf{1} - \bM ) \odot \bd \, ,
\end{aligned}
\end{equation}
in which $y_{t'_i}$ is the original label of $x_i$. $R_{\text{elastic}}(\cdot)$ and $R_{\text{smooth}}(\cdot)$ indicates the elastic-net and the smoothness regularization. 

\begin{table}[h!]
    \centering
\begin{tabular}{|c|c|c|}
\hline
{\tt Neural Cleanse}                    & $\lambda_1$ & \begin{tabular}[c]{@{}c@{}}$1\times e^{-1}$, $1\times e^{-3}$, \\$1\times e^{-5}$, $1\times e^{-7}$\end{tabular} \\ \hline
\multirow{6}{*}{\sys} & $\lambda_1$     & $1\times e^{-3}$, $1\times e^{-5}$                                                        \\ \cline{2-3} 
                     & $\lambda_2$      & $1\times e^{-4}$, $1\times e^{-6}$                                                        \\ \cline{2-3} 
                     & $\lambda_3$      & $1\times e^{-5}$, $1\times e^{-7}$                                                        \\ \cline{2-3} 
                     & $\lambda_4$      & $1\times e^{-6}$, $1\times e^{-8}$                                                        \\ \cline{2-3} 
                     & $\lambda_5$      & $1\times e^{-4}$, $1\times e^{-6}$                                                        \\ \cline{2-3} 
                     & $\lambda_6$      & $1\times e^{-2}$, $1\times e^{-4}$                                                        \\ \hline
\end{tabular}
\caption{The values of each hyperparameters in hyperparameters sensitivity experiments.}
\label{tab:random_used}
\vspace{-10pt}
\end{table}

\vspace{-20pt}
\section{Hyperparameters}
Here Table~\ref{tab:hyper_used} are all the hyperparameters of \sys we used in our experiments. It should be noted ImageNet and LFW share the same set of hyperparameters. We also exhibit all the selected values of each hyperparameters used in the hyperparameter sensitivity experiments in Table~\ref{tab:random_used}.

\begin{table}[hp]
    \centering
    \begin{tabular}{|c|c|c|}
    \hline
    Hyperparameters & GTSRB & ImageNet \& LFW \\
    \hline
    \#Epochs & 500 & 500 \\
    \hline
    Learning Rate & $1\times e^{-3}$ & $1\times e^{-3}$ \\
    \hline
    Optimizor & {\tt Adam} & {\tt Adam}\\
    \hline
    Mask Regularizor & elastic net & elastic net \\
    \hline
    Delta Regularizor & elasetic net & $l_1$ \\
    \hline
    $\lambda_1$ & $1\times e^{-6}$ & $1\times e^{-3}$ \\
    \hline
    $\lambda_2$ & $1\times e^{-5}$ & $1\times e^{-5}$\\
    \hline
    $\lambda_3$ & $1\times e^{-7}$ & $1\times e^{-6}$ \\
    \hline
    $\lambda_4$ & $1\times e^{-8}$ & $1\times e^{-9}$ \\
    \hline
    $\lambda_5$ & $1\times e^{-6}$ & $1\times e^{-6}$ \\
    \hline
    $\lambda_6$ & $1\times e^{-2}$ & $1\times e^{-2}$ \\
    \hline
    \end{tabular}
    \caption{The values of the hyperparameters used in our experiments.}
    \label{tab:hyper_used}
\vspace{-10pt}
\end{table}

\vspace{-20pt}
\section{Model Architectures}
\label{app:model}
There are 4 different model architectures mentioned in this paper, 6 Conv + 2 MaxPooling CNN, 10 Conv + 5 MaxPooling CNN, VGG16~\cite{simonyan2014very} and ResNet50~\cite{he2016deep}. The architecture of 6 Conv + 2 MaxPooling CNN, 10 Conv + 5 MaxPooling CNN are shown in Figure~\ref{fig:6conv} and Figure~\ref{fig:10conv} respectively. The convolutional layers in these two networks have kernels with size $3\times3$ and the pool size of MaxPooling layers is $2\times2$. Figure~\ref{fig:vgg16} shows the architecture of VGG16. We refer interested readers to \cite{simonyan2014very} for detailed information. The architecture of ResNet50 is too large so we do not include the complete structure here. We refer interested readers to this paper \cite{he2016deep} for a detailed description and diagram of ResNet50.

\section{Visual Results}
\begin{figure*}[htbp]
\centering
\includegraphics[width=0.95\textwidth]{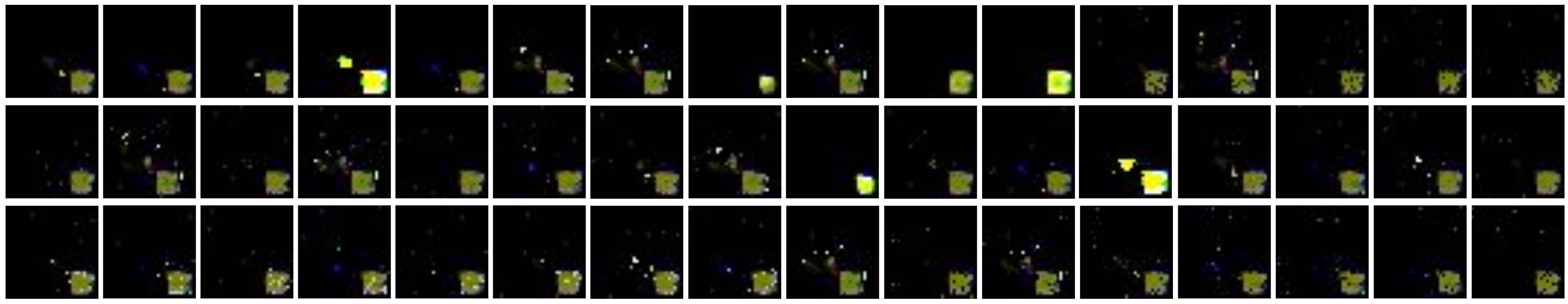}
\caption{The rest results of hyperparameter sensitivity experiments.}
\label{fig:sen_rest}
\end{figure*}

We first show the rest of the results in hyperparameter sensitivity experiments in Figure~\ref{fig:sen_rest}. We then include all the originally injected triggers and the triggers detected in the target labels using {\tt Neural Cleanse} or \sys in Table \ref{tab:res_lfw}, \ref{tab:res_gtsrb}, and \ref{tab:res_imagenet}.
\tikzstyle{conv_layer} = [rectangle, minimum width=3cm, minimum height=0.2cm, rotate=90, text centered, draw=black, fill=white!30]

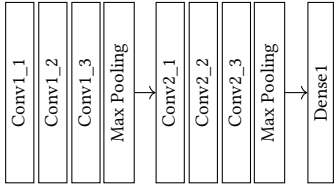
\begin{figure}
    \centering
     \begin{tikzpicture}[node distance=0.55cm, every node/.style={scale=0.8}]
     \node (conv1_1) [conv_layer] {Conv1\_1};
     \node (conv1_2) [conv_layer, below of = conv1_1] {Conv1\_2};
     \node (conv1_3) [conv_layer, below of = conv1_2] {Conv1\_3};
     \node (pool1) [conv_layer, below of = conv1_3] {Max Pooling};
     
     \node (conv2_1) [conv_layer, below of = pool1, yshift=-0.3cm] {Conv2\_1};
     \node (conv2_2) [conv_layer, below of = conv2_1] {Conv2\_2};
     \node (conv2_3) [conv_layer, below of = conv2_2] {Conv2\_3};
     \node (pool2) [conv_layer, below of = conv2_3] {Max Pooling};
     
     \node (dense1) [conv_layer, below of = pool2, yshift=-0.3cm] {Dense1};
     
     \draw [->] (pool1) -- (conv2_1);
     \draw [->] (pool2) -- (dense1);
     \end{tikzpicture}
    \caption{Model Architecture of 6 Conv + 2 MaxPooling Convolutional Neural Network.}
    \label{fig:6conv}
    \vspace{-5pt}
\end{figure}

\begin{figure}
    \centering
      \begin{tikzpicture}[node distance=0.55cm, every node/.style={scale=0.8}]
     \node (conv1_1) [conv_layer] {Conv1\_1};
     \node (conv1_2) [conv_layer, below of = conv1_1] {Conv1\_2};
     \node (pool1) [conv_layer, below of = conv1_2] {Max Pooling};
     
     \node (conv2_1) [conv_layer, below of = pool1, yshift=-0.3cm] {Conv2\_1};
     \node (conv2_2) [conv_layer, below of = conv2_1] {Conv2\_2};
     \node (pool2) [conv_layer, below of = conv2_2] {Max Pooling};

     \node (conv3_1) [conv_layer, below of = pool2, yshift=-0.3cm] {Conv3\_1};
     \node (conv3_2) [conv_layer, below of = conv3_1] {Conv3\_2};
     \node (pool3) [conv_layer, below of = conv3_2] {Max Pooling};
     
     \node (conv4_1) [conv_layer, below of = pool3, yshift=-0.3cm] {Conv4\_1};
     \node (conv4_2) [conv_layer, below of = conv4_1] {Conv4\_2};
     \node (pool4) [conv_layer, below of = conv4_2] {Max Pooling};
     
     \node (conv5_1) [conv_layer, below of = pool4, yshift=-0.3cm] {Conv5\_1};
     \node (conv5_2) [conv_layer, below of = conv5_1] {Conv5\_2};
     \node (pool5) [conv_layer, below of = conv5_2] {Max Pooling};
     
     \node (dense1) [conv_layer, below of = pool5, yshift=-0.3cm] {Dense1};
     
     \draw [->] (pool1) -- (conv2_1);
     \draw [->] (pool2) -- (conv3_1);
     \draw [->] (pool3) -- (conv4_1);
     \draw [->] (pool4) -- (conv5_1);
     \draw [->] (pool5) -- (dense1);
     \end{tikzpicture}
    \caption{Model Architecture of 10 Conv + 5 MaxPooling Convolutional Neural Network.}
    \label{fig:10conv}
    \vspace{-5pt}
\end{figure}

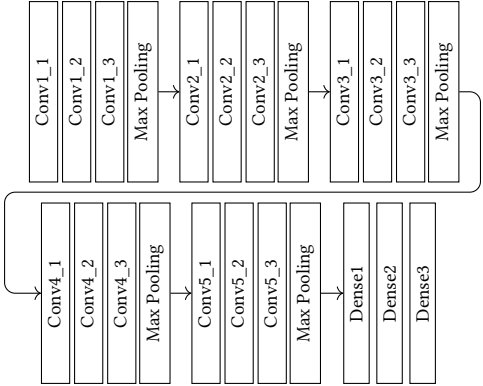
\begin{figure}
    \centering
     \begin{tikzpicture}[node distance=0.55cm, every node/.style={scale=0.8}]
     \node (conv1_1) [conv_layer] {Conv1\_1};
     \node (conv1_2) [conv_layer, below of = conv1_1] {Conv1\_2};
     \node (conv1_3) [conv_layer, below of = conv1_2] {Conv1\_3};
     \node (pool1) [conv_layer, below of = conv1_3] {Max Pooling};
     
     \node (conv2_1) [conv_layer, below of = pool1, yshift=-0.3cm] {Conv2\_1};
     \node (conv2_2) [conv_layer, below of = conv2_1] {Conv2\_2};
     \node (conv2_3) [conv_layer, below of = conv2_2] {Conv2\_3};
     \node (pool2) [conv_layer, below of = conv2_3] {Max Pooling};
     
     \node (conv3_1) [conv_layer, below of = pool2, yshift=-0.3cm] {Conv3\_1};
     \node (conv3_2) [conv_layer, below of = conv3_1] {Conv3\_2};
     \node (conv3_3) [conv_layer, below of = conv3_2] {Conv3\_3};
     \node (pool3) [conv_layer, below of = conv3_3] {Max Pooling};
     
     \node (conv4_1) [conv_layer, left of = conv1_1, xshift=-2.8cm, yshift=-0.2cm] {Conv4\_1};
     \node (conv4_2) [conv_layer, below of = conv4_1] {Conv4\_2};
     \node (conv4_3) [conv_layer, below of = conv4_2] {Conv4\_3};
     \node (pool4) [conv_layer, below of = conv4_3] {Max Pooling};
     
     \node (conv5_1) [conv_layer, below of = pool4, yshift=-0.3cm] {Conv5\_1};
     \node (conv5_2) [conv_layer, below of = conv5_1] {Conv5\_2};
     \node (conv5_3) [conv_layer, below of = conv5_2] {Conv5\_3};
     \node (pool5) [conv_layer, below of = conv5_3] {Max Pooling};
     
     \node (dense1) [conv_layer, below of = pool5, yshift=-0.3cm] {Dense1};
     \node (dense2) [conv_layer, below of = dense1] {Dense2};
     \node (dense3) [conv_layer, below of = dense2] {Dense3};
     
    \draw [->] (pool1) -- (conv2_1);
    \draw [->] (pool2) -- (conv3_1);
    \draw [rounded corners, ->] (pool3.south) -- ++(8pt,0pt) |- ++(-180pt, -38pt) |- (conv4_1.north);
    \draw [->] (pool4) -- (conv5_1);
    \draw [->] (pool5) -- (dense1);
    \end{tikzpicture}\\

    \caption{Model Architecture of VGG16.}
    \vspace{-10pt}
    \label{fig:vgg16}
\end{figure}
\begin{table}[htbp]
    \centering
    \begin{tabular}{|>{\columncolor{gray}}c|>{\columncolor{gray}}c|>{\columncolor{gray}}c|}
    \hline
    \multicolumn{1}{|>{\columncolor{white}}c|}{Original trigger} & \multicolumn{1}{>{\columncolor{white}}c|}{\tt Neural Cleanse} & \multicolumn{1}{>{\columncolor{white}}c|}{\sys} \\
    \hline
    \includegraphics[width=0.05\textwidth]{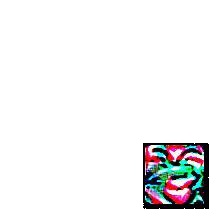} & \includegraphics[width=0.05\textwidth]{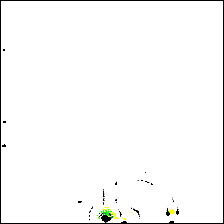} & \includegraphics[width=0.05\textwidth]{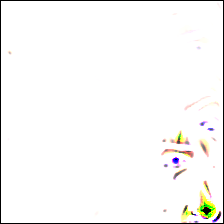}\\
    \hline
    \end{tabular}
    \caption{Original triggers and restored triggers on LFW.}
    \label{tab:res_lfw}
\end{table}

\begin{table*}[htbp]
    \centering
    \setlength{\tabcolsep}{5mm}{
    \begin{tabular}{|c|c|c|c|c|c|c|c|}
    \hline
    Shape & Position & & $6\times6$ & $8\times8$ & $10\times10$ & $12\times12$ & $14\times14$\\
    \hline
    \multirow{6}{*}{Square} & \multirow{3}{*}{\begin{tabular}[c]{@{}c@{}}Top\\ Left\end{tabular}} & \begin{tabular}[c]{@{}c@{}}Original\\ trigger\end{tabular} &  \includegraphics[width=0.05\textwidth]{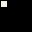} &
    \includegraphics[width=0.05\textwidth]{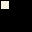} & \includegraphics[width=0.05\textwidth]{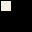} & \includegraphics[width=0.05\textwidth]{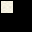} & \includegraphics[width=0.05\textwidth]{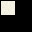}\\
    \cline{3-8}
    & & {\tt Neural Cleanse} & \includegraphics[width=0.05\textwidth]{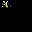} & \includegraphics[width=0.05\textwidth]{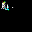} & \includegraphics[width=0.05\textwidth]{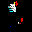} & \includegraphics[width=0.05\textwidth]{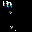} & \includegraphics[width=0.05\textwidth]{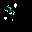}\\
    \cline{3-8}
    & & \sys  & \includegraphics[width=0.05\textwidth]{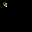} & \includegraphics[width=0.05\textwidth]{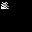} & \includegraphics[width=0.05\textwidth]{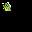} & \includegraphics[width=0.05\textwidth]{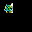} & \includegraphics[width=0.05\textwidth]{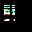}\\
    \cline{2-8}
    & \multirow{3}{*}{\begin{tabular}[c]{@{}c@{}}Bottom\\ Right\end{tabular}} & \begin{tabular}[c]{@{}c@{}}Original\\ trigger\end{tabular} & \includegraphics[width=0.05\textwidth]{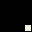} & \includegraphics[width=0.05\textwidth]{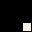} & \includegraphics[width=0.05\textwidth]{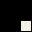} & \includegraphics[width=0.05\textwidth]{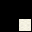} & \includegraphics[width=0.05\textwidth]{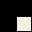}\\
    \cline{3-8}
    & & {\tt Neural Cleanse} & \includegraphics[width=0.05\textwidth]{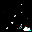} & \includegraphics[width=0.05\textwidth]{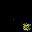} & \includegraphics[width=0.05\textwidth]{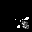} & \includegraphics[width=0.05\textwidth]{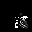} & \includegraphics[width=0.05\textwidth]{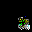}\\
    \cline{3-8}
    & & \sys & \includegraphics[width=0.05\textwidth]{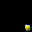} & \includegraphics[width=0.05\textwidth]{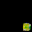} & \includegraphics[width=0.05\textwidth]{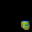} & \includegraphics[width=0.05\textwidth]{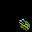} & \includegraphics[width=0.05\textwidth]{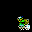}\\
    \hline
    \multirow{6}{*}{Firefox} & \multirow{3}{*}{\begin{tabular}[c]{@{}c@{}}Top\\ Right\end{tabular}} & \begin{tabular}[c]{@{}c@{}}Original\\ trigger\end{tabular} & \includegraphics[width=0.05\textwidth]{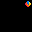} & \includegraphics[width=0.05\textwidth]{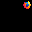} & \includegraphics[width=0.05\textwidth]{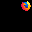} & \includegraphics[width=0.05\textwidth]{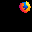} & \includegraphics[width=0.05\textwidth]{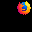} \\
    \cline{3-8}
    & & {\tt Neural Cleanse} & \includegraphics[width=0.05\textwidth]{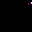} & \includegraphics[width=0.05\textwidth]{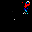} & \includegraphics[width=0.05\textwidth]{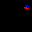} & \includegraphics[width=0.05\textwidth]{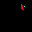} & \includegraphics[width=0.05\textwidth]{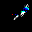}  \\
    \cline{3-8}
    & & \sys & \includegraphics[width=0.05\textwidth]{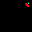} & \includegraphics[width=0.05\textwidth]{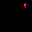} & \includegraphics[width=0.05\textwidth]{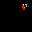} & \includegraphics[width=0.05\textwidth]{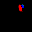} & \includegraphics[width=0.05\textwidth]{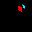}\\
    \cline{2-8}
    & \multirow{3}{*}{\begin{tabular}[c]{@{}c@{}}Bottom\\ Left\end{tabular}} & \begin{tabular}[c]{@{}c@{}}Original\\ trigger\end{tabular} & \includegraphics[width=0.05\textwidth]{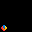} & \includegraphics[width=0.05\textwidth]{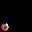} & \includegraphics[width=0.05\textwidth]{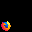} & \includegraphics[width=0.05\textwidth]{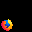} & \includegraphics[width=0.05\textwidth]{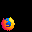}\\
    \cline{3-8}
    & & {\tt Neural Cleanse} & \includegraphics[width=0.05\textwidth]{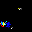} & \includegraphics[width=0.05\textwidth]{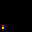} & \includegraphics[width=0.05\textwidth]{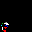} & \includegraphics[width=0.05\textwidth]{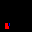} & \includegraphics[width=0.05\textwidth]{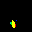} \\
    \cline{3-8}
    & & \sys & \includegraphics[width=0.05\textwidth]{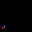} & \includegraphics[width=0.05\textwidth]{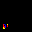} & \includegraphics[width=0.05\textwidth]{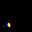} & \includegraphics[width=0.05\textwidth]{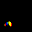} & \includegraphics[width=0.05\textwidth]{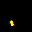}\\
    \hline
    \end{tabular}}
    \caption{Original triggers and restored triggers on GTSRB.}
    \label{tab:res_gtsrb}
\end{table*}
\begin{table*}
    \centering
    \setlength{\tabcolsep}{5mm}{
    \begin{tabular}{|c|c|c|>{\columncolor{gray}}c|>{\columncolor{gray}}c|>{\columncolor{gray}}c|>{\columncolor{gray}}c|>{\columncolor{gray}}c|}
    \hline
    Shape & Position & & \multicolumn{1}{>{\columncolor{white}}c|}{$20\times20$} & \multicolumn{1}{>{\columncolor{white}}c|}{$40\times40$} & \multicolumn{1}{c|}{$60\times60$} & \multicolumn{1}{>{\columncolor{white}}c|}{$80\times80$} & \multicolumn{1}{>{\columncolor{white}}c|}{$100\times100$}\\
    \hline
    \multirow{6}{*}{Square} & \multirow{3}{*}{\begin{tabular}[c]{@{}c@{}}Top\\ Left\end{tabular}} & \begin{tabular}[c]{@{}c@{}}Original\\ trigger\end{tabular} & \includegraphics[width=0.05\textwidth]{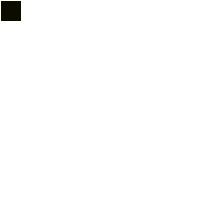} & \includegraphics[width=0.05\textwidth]{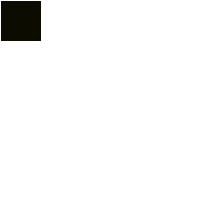} & \includegraphics[width=0.05\textwidth]{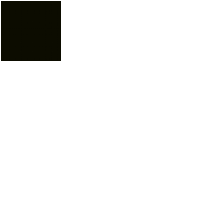} & \includegraphics[width=0.05\textwidth]{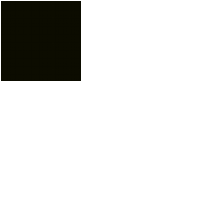} & \includegraphics[width=0.05\textwidth]{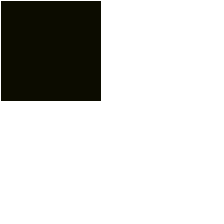}\\
    \cline{3-8}
    & & {\tt Neural Cleanse} & \includegraphics[width=0.05\textwidth]{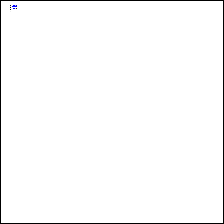} & \includegraphics[width=0.05\textwidth]{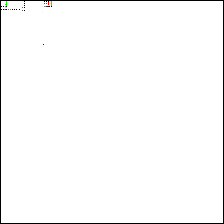} & \includegraphics[width=0.05\textwidth]{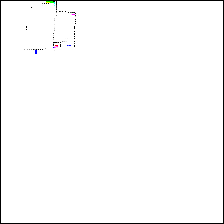} & 
    \includegraphics[width=0.05\textwidth]{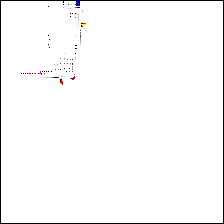} &
    \includegraphics[width=0.05\textwidth]{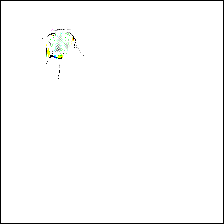}\\
    \cline{3-8}
    & & \sys & \includegraphics[width=0.05\textwidth]{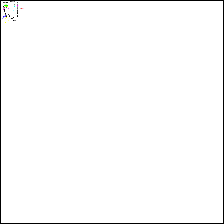} & \includegraphics[width=0.05\textwidth]{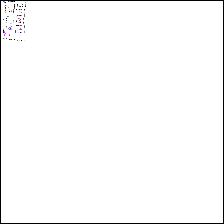} & \includegraphics[width=0.05\textwidth]{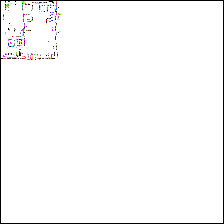} & \includegraphics[width=0.05\textwidth]{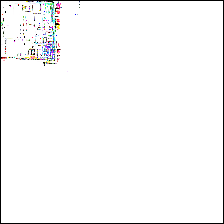} &
    \includegraphics[width=0.05\textwidth]{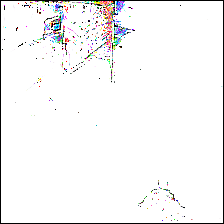}\\
    \cline{2-8}
    & \multirow{3}{*}{\begin{tabular}[c]{@{}c@{}}Bottom\\ Right\end{tabular}} & \begin{tabular}[c]{@{}c@{}}Original\\ trigger\end{tabular} & \includegraphics[width=0.05\textwidth]{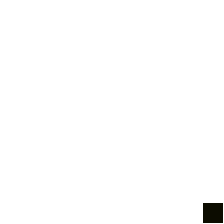} & \includegraphics[width=0.05\textwidth]{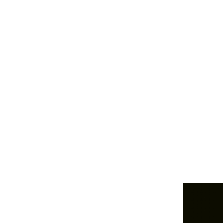} & \includegraphics[width=0.05\textwidth]{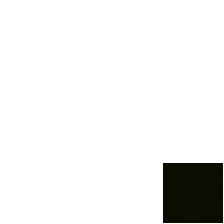} & \includegraphics[width=0.05\textwidth]{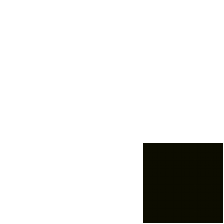} & \includegraphics[width=0.05\textwidth]{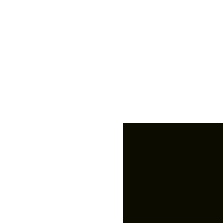} \\
    \cline{3-8}
    & & {\tt Neural Cleanse} & \includegraphics[width=0.05\textwidth]{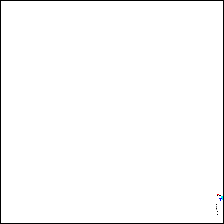} & \includegraphics[width=0.05\textwidth]{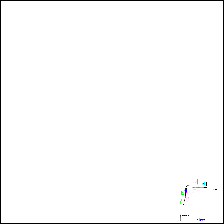} & \includegraphics[width=0.05\textwidth]{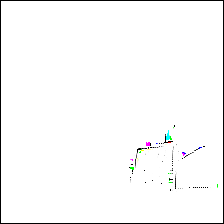} & 
    \includegraphics[width=0.05\textwidth]{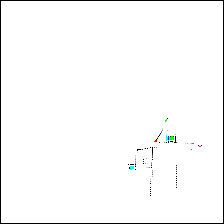} &
    \includegraphics[width=0.05\textwidth]{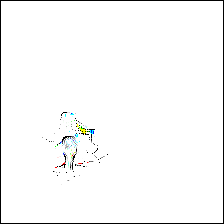}\\
    \cline{3-8}
    & &\sys & \includegraphics[width=0.05\textwidth]{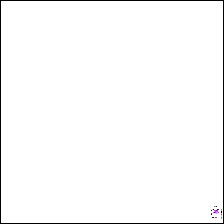} & \includegraphics[width=0.05\textwidth]{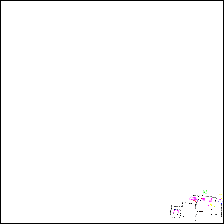} & \includegraphics[width=0.05\textwidth]{img/imagenet/our/br_square_40_fusion_label_1.png} & \includegraphics[width=0.05\textwidth]{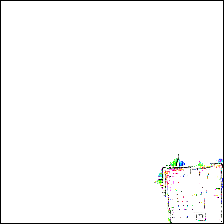} &
    \includegraphics[width=0.05\textwidth]{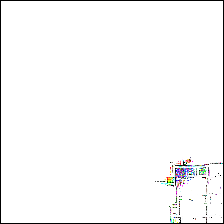}\\
    \hline
    \multirow{6}{*}{Firefox} & \multirow{3}{*}{\begin{tabular}[c]{@{}c@{}}Top\\ Right\end{tabular}} & \begin{tabular}[c]{@{}c@{}}Original\\ trigger\end{tabular} & \includegraphics[width=0.05\textwidth]{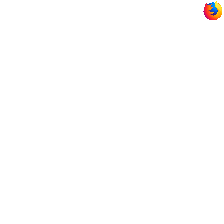} & \includegraphics[width=0.05\textwidth]{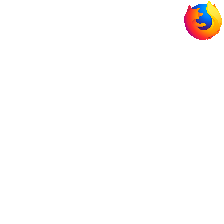} & \includegraphics[width=0.05\textwidth]{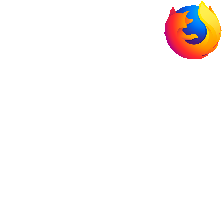} & 
    \includegraphics[width=0.05\textwidth]{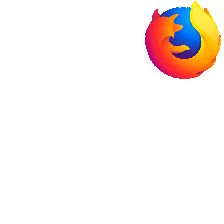} &
    \includegraphics[width=0.05\textwidth]{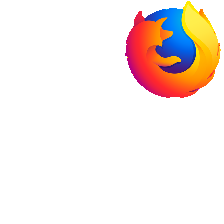} \\
    \cline{3-8}
    & & {\tt Neural Cleanse} & \includegraphics[width=0.05\textwidth]{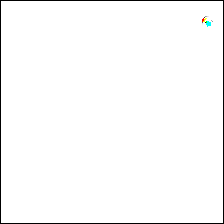} & \includegraphics[width=0.05\textwidth]{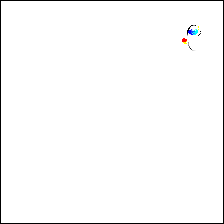} & \includegraphics[width=0.05\textwidth]{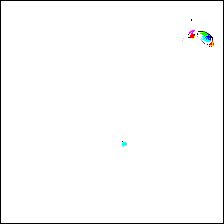} & 
    \includegraphics[width=0.05\textwidth]{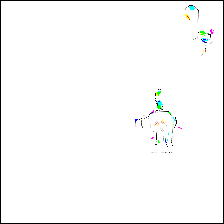} &
    \includegraphics[width=0.05\textwidth]{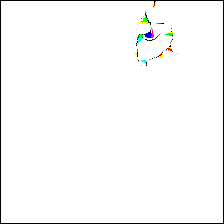} \\
    \cline{3-8}
    & & \sys & \includegraphics[width=0.05\textwidth]{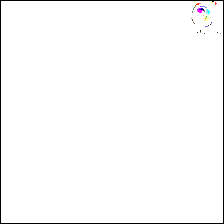} & \includegraphics[width=0.05\textwidth]{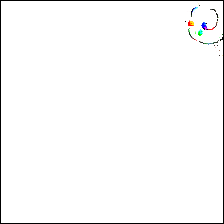} & \includegraphics[width=0.05\textwidth]{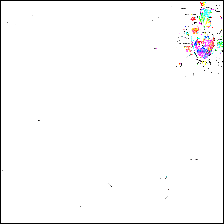} & 
    \includegraphics[width=0.05\textwidth]{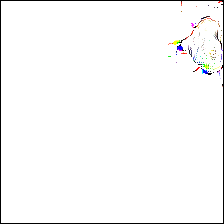} &
    \includegraphics[width=0.05\textwidth]{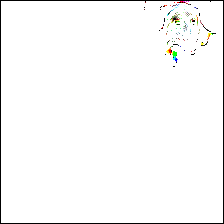} \\
    \cline{2-8}
    & \multirow{3}{*}{\begin{tabular}[c]{@{}c@{}}Bottom\\ Left\end{tabular}} & \begin{tabular}[c]{@{}c@{}}Original\\ trigger\end{tabular} & \includegraphics[width=0.05\textwidth]{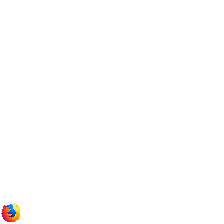} & \includegraphics[width=0.05\textwidth]{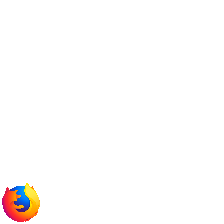} & \includegraphics[width=0.05\textwidth]{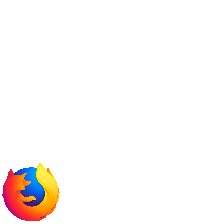} & 
    \includegraphics[width=0.05\textwidth]{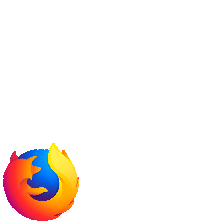} &
    \includegraphics[width=0.05\textwidth]{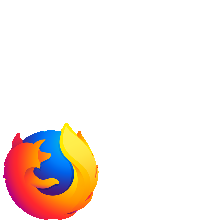} \\
    \cline{3-8}
    & & {\tt Neural Cleanse} & \includegraphics[width=0.05\textwidth]{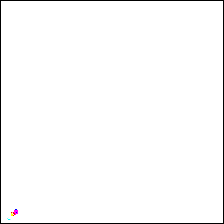} & \includegraphics[width=0.05\textwidth]{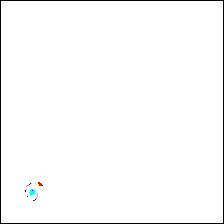} & \includegraphics[width=0.05\textwidth]{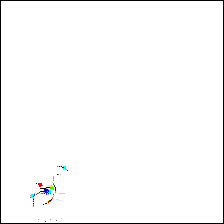} & 
    \includegraphics[width=0.05\textwidth]{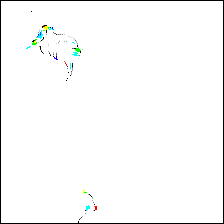} &
    \includegraphics[width=0.05\textwidth]{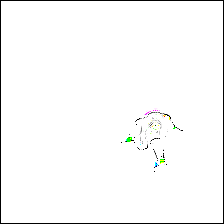} \\
    \cline{3-8}
    & & \sys & \includegraphics[width=0.05\textwidth]{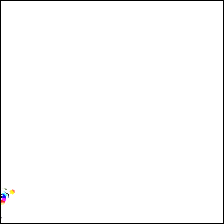} & \includegraphics[width=0.05\textwidth]{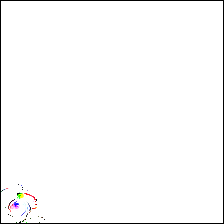} & \includegraphics[width=0.05\textwidth]{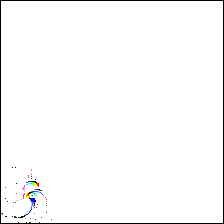} & \includegraphics[width=0.05\textwidth]{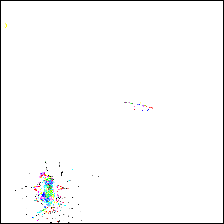} & \includegraphics[width=0.05\textwidth]{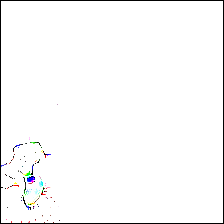}\\
    \hline
    \end{tabular}}
    \caption{Original triggers and restored triggers on ImageNet.}
    \label{tab:res_imagenet}
\end{table*}

\end{document}